\documentclass[twocolumn]{aastex63}
\usepackage{float}
\usepackage{graphicx}
\usepackage{amsmath}
\usepackage{natbib}
\usepackage{color}
\usepackage{verbatim}
\usepackage[utf8]{inputenc}
\usepackage{mathtools}
\usepackage{amssymb}
\usepackage{textcomp}
\usepackage{totcount}
\usepackage{enumitem}
\usepackage{braket}
\usepackage{lipsum}

\newcommand{\um}{\,\textmu m}

\newtotcounter{citnum} 
\def\oldbibitem{} \let\oldbibitem=\bibitem
\def\bibitem{\stepcounter{citnum}\oldbibitem}

\defcitealias{beasley2015}{B15}

\begin{document}
%{\LARGE\total{citnum}}
\title{The Panchromatic Hubble Andromeda Treasury: Triangulum Extended Region (PHATTER). V.\\ The Structure of M33 in Resolved Stellar Populations}

\author[0000-0003-2599-7524]{Adam Smercina}
\affiliation{Department of Astronomy, University of Washington, Box 351580, Seattle, WA 98195-1580, USA}

\author[0000-0002-1264-2006]{Julianne J. Dalcanton}
\affiliation{Center for Computational Astrophysics, Flatiron Institute, 162 Fifth Ave, New York, NY 10010, USA}
\affiliation{Department of Astronomy, University of Washington, Box 351580, Seattle, WA 98195-1580, USA}

\author[0000-0002-7502-0597]{Benjamin F. Williams}
\affiliation{Department of Astronomy, University of Washington, Box 351580, Seattle, WA 98195-1580, USA}

\author[0000-0002-0786-7307]{Meredith J. Durbin}
\affiliation{Department of Astronomy, University of Washington, Box 351580, Seattle, WA 98195-1580, USA}

\author[0000-0002-0786-7307]{Margaret Lazzarini}
\affiliation{Division of Physics, Mathematics, and Astronomy, California Institute of Technology, 1200 E California Blvd., Pasadena, CA 91125, USA}

\author[0000-0002-5564-9873]{Eric F. Bell}
\affiliation{Department of Astronomy, University of 
Michigan, 323 West Hall, 1085 S. University Ave., Ann Arbor, MI, 48105-1107, USA} 

\author[0000-0003-1680-1884]{Yumi Choi}
\affiliation{ NSF's National Optical-Infrared Astronomy Research Laboratory, 950 N. Cherry Avenue, Tucson, AZ 85719, USA}
\affiliation{Department of Astronomy, University of California, Berkeley, CA 94720, USA} 
  
\author[0000-0001-8416-4093]{Andrew Dolphin}
\affiliation{Raytheon Technologies, 1151 E. Hermans Road, Tucson, AZ 85756, USA}
\affiliation{Steward Observatory, University of Arizona, Tucson, AZ 85726, USA}

\author[0000-0003-0394-8377]{Karoline Gilbert}
\affiliation{Space Telescope Science Institute, 3700 San Martin Dr., Baltimore, MD 21218, USA}
\affiliation{The William H. Miller III Department of Physics \& Astronomy, Bloomberg Center for Physics and Astronomy, Johns Hopkins University, 3400 N. Charles Street, Baltimore, MD 21218, USA}

\author[0000-0001-8867-4234]{Puragra Guhathakurta}
\affiliation{UCO/Lick Observatory, Department of Astronomy \& Astrophysics, University of California Santa Cruz, 1156 High Street, Santa Cruz,
California 95064, USA}

\author[0000-0001-9605-780X]{Eric W. Koch}
\affiliation{Center for Astrophysics $\mid$ Harvard \& Smithsonian, 60 Garden Street, Cambridge, MA 02138, USA}

\author[0000-0001-8481-2660]{Amanda C.N. Quirk}
\affiliation{Department of Astronomy, Columbia University, 538 West 120th Street, New York, NY 10027, USA}

\author[0000-0003-4996-9069]{Hans-Walter Rix}
\affiliation{Max-Planck-Institut f\"{u}r Astronomie, K\"{o}nigstuhl 17, D-69117 Heidelberg, Germany}

\author[0000-0002-5204-2259]{Erik Rosolowsky}
\affiliation{Department of Physics, University of Alberta, Edmonton, AB T6G 2E1, Canada}

\author[0000-0003-0248-5470]{Anil Seth}
\affiliation{Department of Physics Astronomy, University of Utah, 115 South 1400 East, Salt Lake City, UT 84112, USA}

\author[0000-0003-0605-8732]{Evan D. Skillman}
\affiliation{Minnesota Institute for Astrophysics, School of Physics and Astronomy, University of  Minnesota, 116 Church St. SE, Minneapolis, MN 55455, USA}

\author[0000-0002-6442-6030]{Daniel R. Weisz}
\affiliation{Department of Astronomy, University of California, Berkeley, CA 94720, USA}

\correspondingauthor{Adam Smercina}
\email{asmerci@uw.edu}

\vspace{-3pt}
\begin{abstract}
    We present a detailed analysis of the the structure of the Local Group flocculent spiral galaxy M33, as measured using the Panchromatic Hubble Andromeda Treasury Triangulum Extended Region (PHATTER) survey. Leveraging the multiwavelength coverage of PHATTER, we find that the oldest populations are dominated by a smooth exponential disk with two distinct spiral arms and a classical central bar --- completely distinct from what is seen in broadband optical imaging, and the first-ever confirmation of a bar in M33. We estimate a bar extent of $\sim$1\,kpc. The two spiral arms are asymmetric in orientation and strength, and likely represent the innermost impact of the recent tidal interaction responsible for M33's warp at larger scales. The flocculent multi-armed morphology for which M33 is known is only visible in the young upper main sequence population, which closely tracks the morphology of the ISM. We investigate the stability of M33's disk, finding $Q\,{\sim}\,1$\ over the majority of the disk. We fit multiple components to the old stellar density distribution and find that, when considering recent stellar kinematics, M33's bulk structure favors the inclusion of an accreted halo component, modeled as a broken power-law. The best-fit halo model has an outer power-law index of $-$3 and accurately describes observational evidence of M33's stellar halo from both resolved stellar spectroscopy in the disk and its stellar populations at large radius. Integrating this profile yields a total halo stellar mass of ${\sim}5{\times}10^8\,M_{\odot}$, giving a total stellar halo mass fraction of 16\%, most of which resides in the innermost 2.5\,kpc.
\end{abstract}

\NewPageAfterKeywords

\section{Introduction}
\label{sec:intro}

Galaxies have long been observed to display a wide variety of morphologies, which classification schemes such as the `Hubble Sequence' (see \citealt{sandage2005} for a review) have attempted to organize in an evolutionary context. It is now relatively well understood that much of this diversity broadly correlates with stellar mass and mass surface density, reflecting both the assembly histories of galaxies and their environments \citep[e.g.,][and references therein]{conselice2014}. Among the different structural classes that galaxies belong to, stellar disks dominate the morphologies of isolated near-$L_{\star}$\ galaxies like the Milky Way (MW), with disks arising naturally in current galaxy formation theory \citep{tinsley&larson1978,mo1998,silk2001,stringer&benson2007}. However, what drives the observed diversity of disk structure within galaxies' disks --- e.g., their scale, spiral arms, stellar bars --- has remained an active area of research. Galaxies like Triangulum (hereafter M33) offer a unique perspective on this problem, as they occupy precisely the mass scale at which galactic disks and spiral structure begin to emerge from the diffuse structure of irregular systems \citep[$M_{\star}\,{\sim}\,10^9\, M_{\odot}$; e.g.,][]{gallagher&hunter1984}. Understanding the origins of disks in these low-mass galaxies is therefore a powerful lens through which to better understand disk formation and evolution at all mass scales. M33 is the nearest prototypical flocculent spiral galaxy, and is therefore a crucial laboratory for understanding disk formation. 

M33 has nearly the same stellar mass ($M_{\star}\,{\sim}\,3{\times}10^9\ M_{\odot}$; \citealt{vandermarel2012}) and metallicity ([Fe/H]\,$\sim$\,$-$0.5; \citealt{beasley2015}) as the Large Magellanic Cloud \citep[LMC;][]{bica1998,carrera2008,nidever2020}. Yet, the LMC's recent interaction with the SMC as they fall in tandem into the MW halo has resulted in a disturbed morphology \citep[e.g.,][]{vandermarel2002,besla2012,choi2018a,choi2018b}, making it difficult to disentangle the secular and interaction-driven origins of its spiral arms and prominent bar. Similarly to the LMC, M33 is the largest satellite galaxy of M31, though much further away from M31 than the LMC to the MW. However, in contrast, it is relatively undisturbed, displaying a flocculent or `pinwheel' spiral structure at visible wavelengths. The few hints at M33's dynamic involvement in the M31 Group are a relatively prominent warp in its outer disk in both gas and stars \citep[e.g.,][]{sandage&humphreys1980,putman2009,mcconnachie2010}, and a corresponding H\,\textsc{i} ``bridge'' between M33 and M31 \citep{wolfe2013,lockman2012}. Recent spectroscopic observations of stars in M33's disk indicate that it also hosts a high-velocity dispersion stellar component, with a high central mass fraction, that may comprise M33's accreted stellar halo \citep{gilbert2022}. However, the shape of this halo component, and its relevance to M33's interaction history, is still unknown. M33's general morphology is shared among all isolated and seemingly-undisturbed LMC-mass galaxies in the Local Volume (LV), such as NGC 300 \citep[e.g.,][]{bland-hawthorn2005,gogarten2010}, NGC 2403 \citep[e.g.,][]{hudon1989,williams2013}, and NGC 2976 \citep[e.g.,][]{williams2010} --- all of which exhibit a comparable flocculent spiral structure, with no notable bulge. 

M33's structure has been studied extensively at different wavelengths, from broadband imaging of stellar tracers in the optical \citep[e.g.,][]{sandage&humphreys1980},  ultraviolet \citep{thilker2005}, and near-infrared \citep[e.g.,][]{regan&vogel1994,jarrett2003}, as well as observations of the dust continuum \citep[e.g.,][]{rice1990,tabatabaei2007} and H\,\textsc{i} 21\,cm emission \citep[e.g.,][]{koch2018}. Unlike its more disturbed counterpart in the Local Group (the LMC), M33 shows little obvious evidence of a central bar, much like its field analogs in the LV. There have, however, been hints that the flocculent spiral structure of M33-analogs may be dependent on the stellar populations probed, as both M33 and NGC 300 exhibit a less flocculent structure in the near-infrared \citep{cepa1988,regan&vogel1994}. Furthermore, in infrared imaging, M33's central isophotes show a weak, bar-shaped component, though it has been argued that this may simply be an inner extension of the spiral arms \citep{regan&vogel1994}. The presence of a bar has been investigated in depth in the intervening years, with some arguing for \citep[e.g.,][]{corbelli&walterbos2007} or against \citep[e.g.,][]{sellwood2019} the presence of a weak central bar. 

\begin{figure*}[t]
\centering
\leavevmode
\includegraphics[width={\linewidth}]{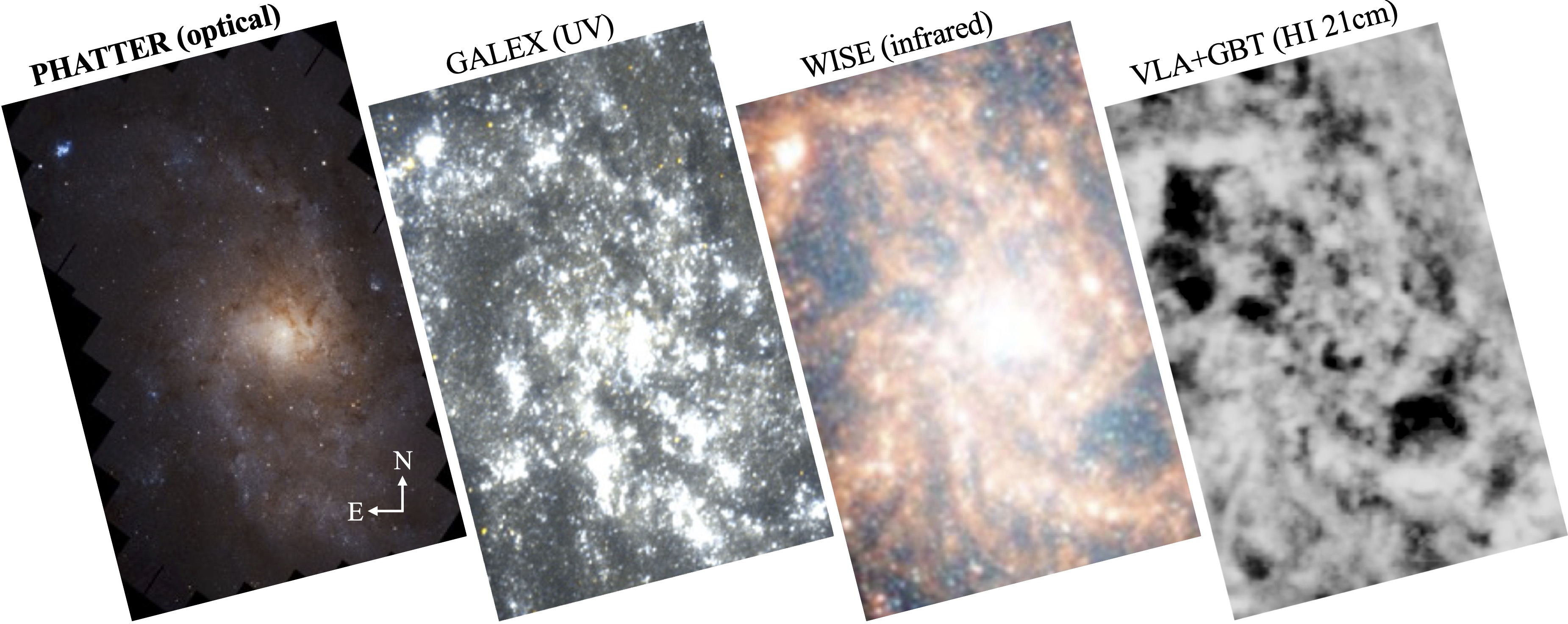}
\caption{M33 imaged at four different wavelengths, each within the PHATTER footprint: optical (PHATTER; \citealt{williams2021}, this work), ultraviolet (GALEX; \citealt{thilker2005}), near-infrared (WISE; \citealt{jarrett2019}), and H\,\textsc{i} 21\,cm (VLT+GBT; \citealt{koch2018}). Images are shown in a typical north-up, east-left orientation; the GALEX, WISE, and H\,\textsc{i} images are tilted to be consistent with the PHATTER footprint. All four images show M33's flocculent spiral structure, though the optical and near-infrared images clearly show two more dominant spiral arms in M33's central regions.}
\label{fig:multi-wave}
\end{figure*}

Though it shares morphological similarity with many isolated analogs in the Local Volume, M33's outskirts betray its role as a satellite in the Local Group environment. It possesses a prominent warp in its outer gas disk \citep[e.g.,][]{sandage&humphreys1980,mcconnachie2006,putman2009,corbelli2014}, as well as a skewed low-surface brightness (SB) stellar substructure extending out to 40 kpc from its center, twisted relative to the disk plane \citep{mcconnachie2010,ibata2014}. Additionally, M33 experienced a disk-wide enhancement in its star formation rate (SFR) $\sim$2\,Gyr ago, roughly coincident with a similar enhancement in M31 \citep{williams2009,bernard2012,williams2015,williams2017}.\footnote{This star-forming `burst' in M31 is now thought to be likely due to a completed merger with a galaxy even more massive than M33 \citep{hammer2018,dsouza&bell2018b,bhattacharya2021}.} M33's disturbed outer morphology, coupled with the similar timescales of the enhancements in both M31's and M33's star formation histories (SFHs), suggests a possible recent close passage between the two. However, orbital modeling based on proper motions cast doubt on this possibility \citep[e.g.,][]{patel2017,vandermarel2019}, which we discuss further in \S\,\ref{sec:interaction}. Given its proximity and signatures of both secular and environmental processes at work, M33 is clearly a unique laboratory for studying the relative importance of secular and interaction-driven processes in shaping the morphology of low-mass spiral galaxies. 

In recent years, results from the Panchromatic Hubble Andromeda Treasury (PHAT) survey \citep{dalcanton2012} have shown the power of multi-band resolved star datasets in extracting the subtle imprints of M31's evolutionary history on its structure \citep{williams2017}, SFH \citep{williams2015,williams2017}, and kinematics \citep{dorman2015}. Recently, the Panchromatic Hubble Andromeda Treasury Triangulum Extended Region (PHATTER) survey has applied this same methodology to resolve 22 million stars in M33, in six photometric filters \citep{williams2021}. 

In this fifth paper using observations from the PHATTER survey, we present a structural analysis of M33, out to 3\,kpc, across several different stellar sub-populations of different characteristic age. The structure of the paper is as follows: we briefly summarize the observations and data quality of the PHATTER survey (\S\,\ref{sec:obs}), then describe the characterization and selection of the different stellar populations used throughout this paper, and then present our results (\S\,\ref{sec:results}). We produce stellar density maps (\S\,\ref{sec:density}) for each population, conduct an analysis of M33's spiral structure (\S\,\ref{sec:spiral}), and decompose the structure of its old stars (\S\,\ref{sec:decomp}). We discuss our results for M33 in the context of other analogous Magellanic spirals (\S\,\ref{sec:context}) and then summarize our conclusions (\S\,\ref{sec:conclusions}). 

Throughout the paper we assume a distance to M33 of 859\,kpc \citep{degrijs2017}, and a corresponding distance modulus of $m{-}M\,{=}\,24.67$. 

\section{Observations and Reduction}
\label{sec:obs}

We investigate the structure of M33 using resolved star catalogs obtained as part of the PHATTER survey\footnote{\href{https://archive.stsci.edu/hlsp/phatter}{https://archive.stsci.edu/hlsp/phatter}}. The survey strategy and reduction procedure for the PHATTER observations is described in-depth in \cite{williams2021} and has been summarized in \cite{lazzarini2022} and \cite{johnson2022}. Here we summarize the reduction and generation of the stellar catalogs used in this paper, but refer the reader to \cite{williams2021} for complete details. 

PHATTER obtained photometry for 22 million stars in six bands, spanning the ultraviolet to the near-infrared: F475W, F814W (ACS), F275W, F336W (WFC3/UVIS), F110W, F160W (WFC3/IR). The observations were completed using a tiling scheme organized into three ``bricks'' of 3$\times$6 WFC3/IR pointings. In total, the survey area tiled M33's inner 13\farcm2$\times$19\farcm8, reaching a (projected) $\sim$2.5\,kpc along the major axis. 

All individual exposures were aligned using the \textit{Gaia} DR2 astrometric solution \citep[following][]{bajaj2017} to create mosaics, which were used to identify and mask bad pixels and cosmic rays. Point spread function (PSF) fitting photometry was carried out on the aligned images using the DOLPHOT software \citep{dolphin2000,dolphin2016}. Statistically significant centroids were detected in the stacked images of all overlapping exposures. The appropriate PSF was then fit to each location to measure total fluxes and corresponding magnitudes in each band for each source. 

\begin{figure*}[t]
\centering
\leavevmode
\includegraphics[width={0.9\linewidth}]{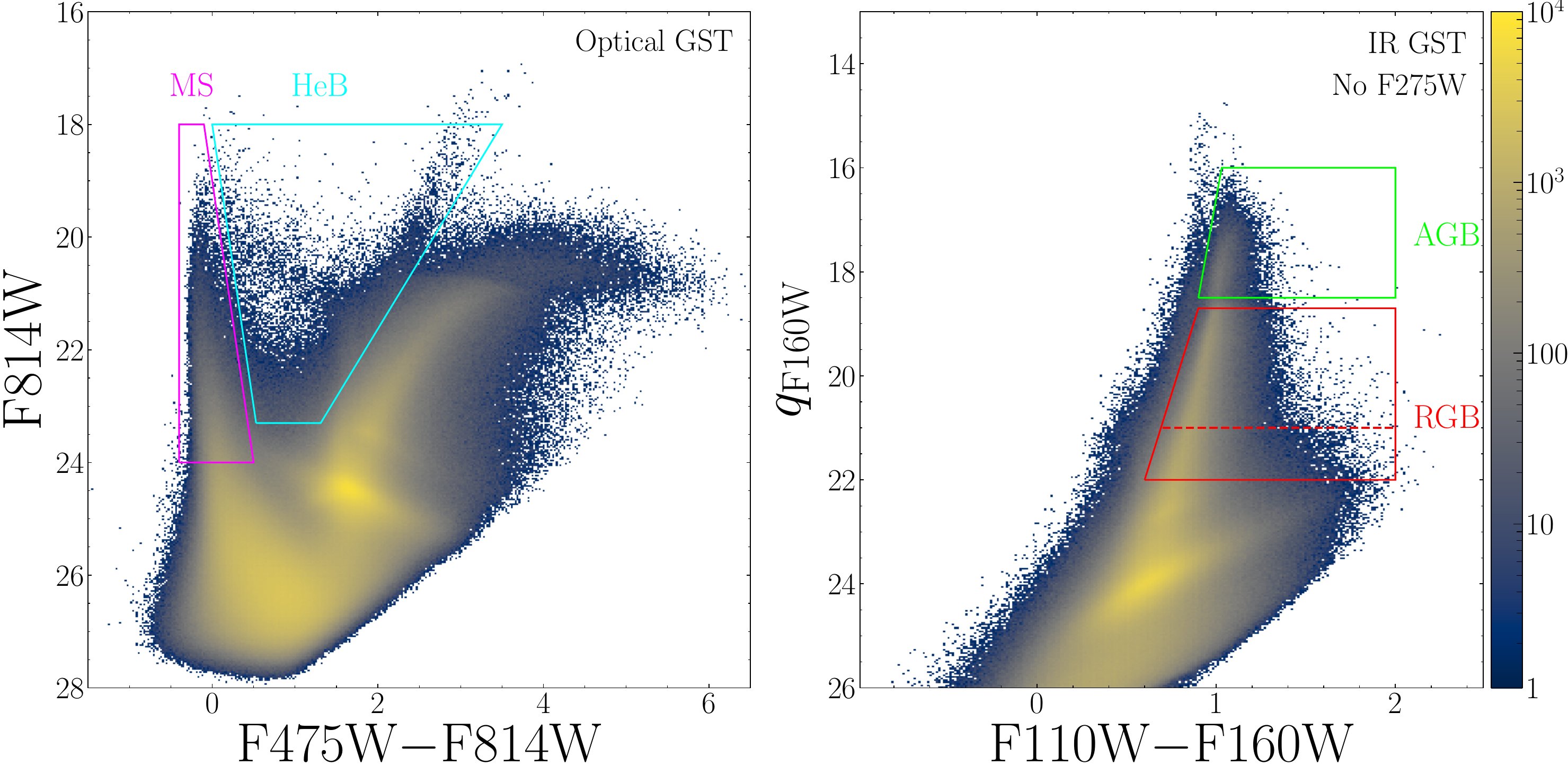}
\caption{Hess diagrams for stars detected in the PHATTER survey. \uline{Left}: F475W$-$F814W vs.\ F814W CMD of stars in PHATTER passing the optical GST criteria. The selection regions for the MS and HeB stars are shown in magenta and cyan, respectively. \uline{Right}: F110W$-$F160W vs.\ $q_{\rm F160W}$\ (reddening-free; defined in Equation \ref{eq:q}) CMD of stars in PHATTER passing the infrared GST criteria, but failing the GST criteria for F275W --- effectively selecting old, red stars. The AGB and RGB selection regions are shown in green and red, respectively. The dashed red line denotes the cutoff magnitude for our shallower selection `inner' RGB selection to account for lower completeness due to crowding.}
\label{fig:selection}
\end{figure*}

Photometric quality is determined from a combination of signal-to-noise (S/N), the central `peakiness' of the point source (\texttt{sharpness}), and how much the source photometry is affected by neighboring sources (\texttt{crowding}). For this paper, we adopt the GST (i.e., `Good Star') photometric quality criteria of \cite{williams2021} that correspond to good measurements of real stars in the PHAT survey. We adopt a uniform condition of S/N\,$>$\,4 for source detections in all bands. For ACS we adopt: \texttt{sharp}$^2$\,$<$\,0.2 and \texttt{crowd}\,$<$\,2.25. For WFC3/UVIS: \texttt{sharp}$^2$\,$<$\,0.15 and \texttt{crowd}\,$<$\,1.30. For WFC3/IR: \texttt{sharp}$^2$\,$<$\,0.15 and \texttt{crowd}\,$<$\,2.25. These parameters have been shown to appropriately balance source completeness, crowding, and contamination, and are discussed in detail in other PHAT survey papers \citep[e.g.,][]{williams2014,williams2021}. 

Artificial star tests (ASTs) were used to assess the accuracy and completeness of the PHATTER photometry. Artificial stellar magnitudes were generated using the MATCH software \citep{dolphin2002}, sampled from the MIST stellar evolution models \citep{choi2016}. PHATTER, similar to PHAT before it, is primarily `crowding-limited' in the optical and near-infrared, where the density of stellar sources in a particular region is the principal limiting factor in the depth and precision with which photometry can be measured for each source. To give a sense of the impact crowding has on source detection: at the lowest stellar densities, we reach 50\% completeness at F814W\,=\,26.8 and F160W\,=\,25.0, while at the highest densities this becomes F814W\,=\,25.2 and F160W\,$\simeq$\,22.9. We use the results of these ASTs to determine the magnitude ranges used in our analysis of stellar populations across M33, such that our selected populations are uniformly complete across the survey footprint.

The final source catalog contains nearly 22 million stars, of which 7.96 million satisfy both the ACS and WFC3/IR GST criteria, while 285,000 of the brightest stars satisfy the GST criteria in all six bands. A color image of the PHATTER survey is shown in Figure \ref{fig:multi-wave}, along with images of M33 within this same field-of-view (FOV) at different wavelengths, for context. Optical and infrared color--magnitude Hess diagrams are shown in Figure \ref{fig:selection} (see \S\,\ref{sec:stpop} for details). 

\newcolumntype{s}{!{\extracolsep{25pt}}c!{\extracolsep{0pt}}}
\newcolumntype{t}{!{\extracolsep{25pt}}l!{\extracolsep{0pt}}}
\newcolumntype{p}{!{\extracolsep{25pt}}r!{\extracolsep{0pt}}}

\begin{deluxetable*}{ttp}[!ht]
\floattable
\tablewidth{0.8\textwidth}
\tablecaption{\textnormal{Stellar Population Selection Criteria}\label{tab:select}}
\tablecolumns{3}
\setlength{\extrarowheight}{3pt}
\tablewidth{\linewidth}
\tabletypesize{\small}
\tablehead{%
\colhead{Type} &
\colhead{Description} &
\colhead{Criteria} \\ \vspace{-3.5mm}
}
\startdata
%\vspace{-2mm} \\
\multicolumn{3}{s}{\textbf{Red Giant Branch Outer}} \\
\hline
Photometry & Quality criteria & F110W\,$<$\,24.5, $q_{\rm F160W}$\,$<$\,22 \\
& & IR\,=\,GST, F275W\,!=\,GST \\
Color--magnitude & Vertices of selection region & \\
& (F110W$-$F160W, $q_{\rm F160W}$) = & (0.6,22),(1.3,22),(1.3,18.7),(0.9,18.7) \\
Radius & Angular distance from center & $d$\,$>$\,0.02 deg \\
\hline
\multicolumn{3}{s}{\textbf{Red Giant Branch Inner}} \\
\hline
Photometry & Quality criteria & F110W\,$<$\,23.5, $q_{\rm F160W}$\,$<$\,21 \\
& & IR\,=\,GST, F275W\,!=\,GST \\
Color--magnitude & Vertices of selection region & \\
& (F110W$-$F160W, $q_{\rm F160W}$) = & (0.6,22),(1.3,22),(1.3,18.7),(0.9,18.7) \\
Radius & Angular distance from center & $d$\,$\leqslant$\,0.02 deg \\
\hline
\multicolumn{3}{s}{\textbf{Asymptotic Giant Branch}} \\
\hline
Photometry & Quality criteria & IR\,=\,GST, F275W\,!=\,GST \\
Color--magnitude & Vertices of selection region & \\
& (F110W$-$F160W, $q_{\rm F160W}$) = & 0.9,18.5),(2,18.5),(2,16),(1.03,16) \\
\hline
\multicolumn{3}{s}{\textbf{Helium Burning Sequence}} \\
\hline
Photometry & Quality criteria & Optical\,=\,GST, IR\,=\,GST \\
Color--magnitude & Vertices of selection region & \\
& (F475W$-$F814W, F814W) = & (0.53,23.3),(1.31,23.3),(2.9,18),(0.0,18) \\
\hline
\multicolumn{3}{s}{\textbf{Main Sequence}} \\
\hline
Photometry & Quality criteria & Optical\,=\,GST, IR\,=\,GST \\
Color--magnitude & Vertices of selection region & \\
& (F475W$-$F814W, F814W) = & ($-$0.4,24),(0.5,24),($-$0.1,18),($-$0.4,18) \\
\enddata
\tablecomments{Criteria for selecting stars belonging to each of the four stellar sub-populations discussed in this paper: RGB, AGB, HeB, and MS.}
\end{deluxetable*}

\section{Stellar Populations Selection}
\label{sec:stpop}
Owing to its six-filter coverage and photometric depth, the PHATTER survey is sensitive to a number of different stellar populations in M33. M33 is actively star-forming across its entire disk \citep[e.g.,][]{boquien2015}, with an average SFR of $\sim$0.74 $M_{\odot}$\,yr$^{-1}$\ over the past 100\,Myr \citep{lazzarini2022} and a complex star formation history (SFH) over the past several Gyrs \citep[e.g.,][]{williams2009}. The color--magnitude diagrams (CMDs) of M33's resolved stellar populations, shown in Figure \ref{fig:selection}, reflect this history of star formation, with different regions tracing stars formed in different epochs of M33's evolution. 

In this paper, we consider four broad stellar sub-populations, tracing different epochs of M33's SFH: the Main Sequence (MS), the core Helium-burning (HeB) phase, the Asymptotic Giant Branch (AGB), and the Red Giant Branch (RGB). We first describe the selection criteria for each of these populations (\S\,\ref{sec:select}), and then estimate the approximate age distributions for each using artificial stellar catalogs sampled from stellar models with the MATCH software (\S\,\ref{sec:fakepop}).

\subsection{Description of Populations and Selection Criteria}
\label{sec:select}

Each of the four distinct populations we consider in this paper were selected by a combination of photometric quality and color--magnitude criteria. We give the criteria for each population in Table \ref{tab:select}. The boundaries of each region in Figure \ref{fig:selection} were chosen visually, based on clear features in the CMDs. We examine the validity of these selection regions in \S\,\ref{sec:fakepop} using synthetic stellar populations. 

First, the stellar main sequence (MS) traces stars currently burning hydrogen in their cores. At the depths achieved by PHATTER, the MS is primarily comprised of young, massive stars. The MS is most separable from other regions of the CMD in the optical filters, which we use for selection. We require selected MS stars to pass GST criteria in all optical and infrared bands to ensure high quality detections. The MS selection region is shown in Figure \ref{fig:selection}.

Next, core Helium-burning are intermediate-mass post-main sequence stars (2--15 $M_{\odot}$) that are fusing Helium in their cores for a brief time \citep[see e.g.,][]{dohm-palmer&skillman2002,weisz2008,mcquinn2011}. These HeB stars populate a `blue' (BHeB) and a `red' (RHeB) sequence, looping between the two throughout this phase, with stars of higher masses occupying higher luminosities on the sequences. HeB stars are considerably brighter than MS stars of the same age, meaning that at similar brightness, they probe somewhat older populations than the MS. Like the MS, we require selected HeB stars to pass GST criteria in all optical and infrared bands. Our HeB selection is done in the optical to maximize the separation from the RGB and MS. The selection region was chosen to encompass both the blue and red tracks, and the faint-end of the selection region is somewhat brighter than the MS to avoid contamination from older stars near the horizontal branch. The HeB selection region is shown in Figure \ref{fig:selection}.

The AGB is the post-main sequence Hydrogen and Helium shell-burning phase of low- to intermediate-mass stars ($\sim$0.8--8 $M_{\odot}$) --- older as a population, on average, than the HeB stars. AGB stars are quite cool and are therefore quite red in their spectral energy distributions. We therefore use the infrared filters, F110W and F160W, for our selection on the CMD \citep[see also][]{dalcanton2012b}. To select intrinsically red stars in the near-infrared CMD, and separate them from stars heavily reddened from \textit{in situ} dust extinction, we transform F160W to a reddening-free magnitude, `$q$', which causes reddened stars to move redward on the CMD, but not to fainter magnitudes. Originally described in \cite{dalcanton2015}\footnote{Inspired by the $Q$\ parameter described in \cite{johnson&morgan1953}.}, `$q$' is defined as follows:
\begin{equation}
\begin{split}
q_{\rm F160W} \equiv {\rm F160W} - \big(({\rm F110W - F160W}) - c_0\big) \\
\times \frac{A_{\rm F160W}/A_V}{A_{\rm F110W}/A_V - A_{\rm F160W}/A_V}, 
\end{split}
\label{eq:q}
\end{equation}
where we adopt $A_{\rm F110W}/A_V$ = 0.3266 and $A_{\rm F160W}/A_V$ = 0.2029 (following \citealt{dalcanton2015}). $c_0$\ is an arbitrary color for which $q_{\rm F160W}$\,$\equiv$\,F160W; we adopt $c_0 = 1.0$. We require selected AGB stars to pass optical and infrared GST criteria, but not UV GST criteria. AGB stars are typically faint in the UV, unlike HeB stars, which retain substantial UV flux despite being redder than massive MS stars. This difference in UV brightness is useful for separating the AGB (and RGB) track from the RHeB track, as they are otherwise adjacent to one another on the CMD, causing cross-contamination in any CMD-only selection. We therefore specifically require selected `old' stars (including both the AGB and RGB) to fail the GST criteria for the FUV filter, F275W. The AGB selection region is shown in Figure \ref{fig:selection}.

Finally, the RGB is populated by evolved stars with masses $\lesssim$\,2 $M_{\odot}$\ in a long-lived Hydrogen shell-burning phase. The Tip of the RGB (TRGB) is an abrupt feature in the luminosity function of red stars in the CMD, signaling the ignition of core Helium fusion and precipitating a sharp drop in luminosity to the Horizontal Branch. As the RGB is a long-lived phase, populated by largely old, low-mass stars that are many magnitudes brighter than their MS counterparts, stars in this region of the CMD are excellent tracers of the bulk of the stellar mass for a given stellar population. This also means that RGB stars are the most numerous of these four selected populations, and therefore most impacted by crowding --- particularly in M33's center. We therefore use an `outer' and `inner' RGB selection: at circular radii 1\farcm2 or less from M33's center, we limit RGB selection to 1\,mag brighter in F110W and $q_{\rm F160W}$\ than for stars with radii greater than 1\farcm2. This radius selection was based on the AST results presented in \cite{williams2021}, where the increased stellar density results in a large change in photometric completeness (the 80\% completeness limit in F160W is $\sim$22.0\,mag at the boundary between these regions). As with the AGB, we select RGB stars in the near-infrared, using the reddening-free $q_{\rm F160W}$, and require that selected RGB stars pass the optical and infrared GST criteria, but fail the F275W GST criteria (to reduce RHeB contamination). The RGB selection region is shown in Figure \ref{fig:selection}.

\begin{figure*}[!ht]
\centering
\leavevmode
\includegraphics[width={\linewidth}]{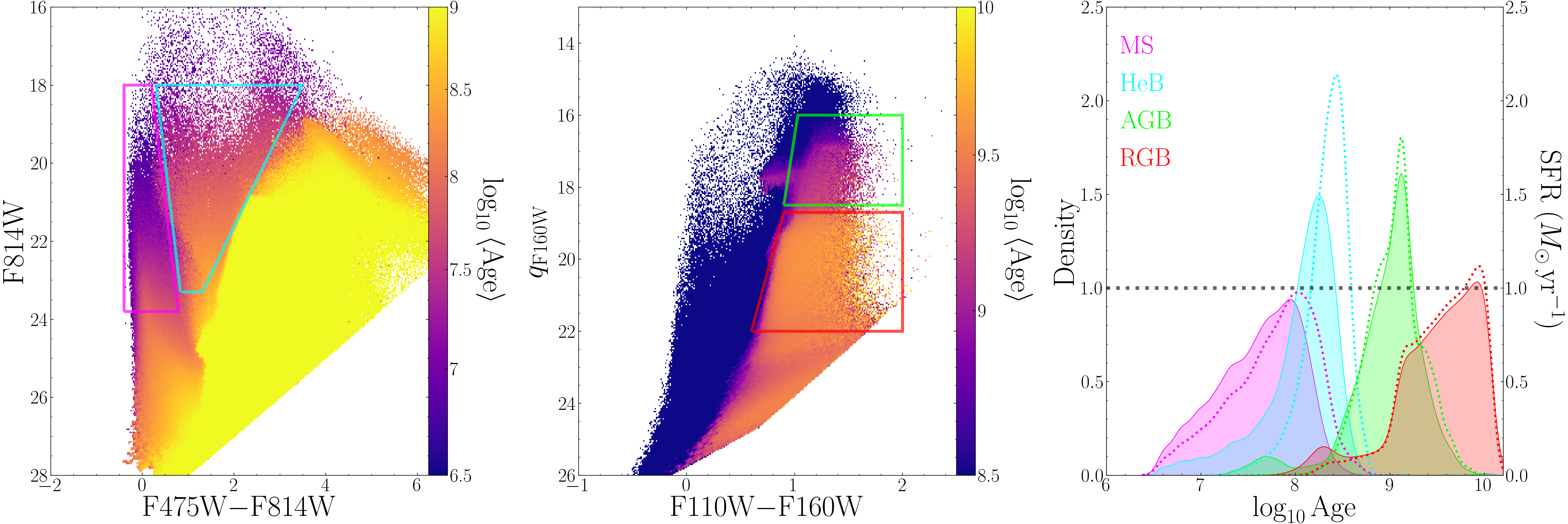}
\caption{Results of our artificial stellar catalog generated with MATCH, assuming a constant SFH, an M33-like age--metallicity relation and distance modulus, and a dust distribution motivated by the CMD analysis of \citet{lazzarini2022}. \textit{Left}: Hess CMD in the F475W and F814W filters of MATCH-generated fake stars, color-coded by the weighted average stellar age in each color--magnitude bin. The MS and HeB selection regions used on the PHATTER catalogs are overlaid. Ages have been scaled to better show differences for younger stars (ages between 3\,Myr and 1\,Gyr). \textit{Center}: Same as the right panel, but for the F110W and F160W filters, and with the AGB and RGB selection regions overlaid. Ages in this panel have been scaled to show differences for older stars (ages between 300\,Myr and 10\,Gyr). \textit{Right}: KDE of the distributions of stellar ages for each population, using a 0.15\,dex smoothing bandwidth. The dotted curves for each show the result for an unreddened synthetic CMD. The four populations --- MS, HeB, AGB, and RGB --- each probe distinct age ranges, with median ages $\sim$0.5\,dex apart, though with substantial overlap.}
\label{fig:fakestars}
\end{figure*}

\subsection{Population Age Distributions}
\label{sec:fakepop}
As described in \S\,\ref{sec:select}, we can say \textit{a priori} that, for the selection criteria given in Table \ref{tab:select} and shown in Figure \ref{fig:selection}, the MS stars generally trace the youngest populations, followed by the HeB stars, then the AGB, and lastly the RGB. However, our selections are very broad, encompassing large ranges of stellar age within each population so as to maximize the number of stars available to explore M33's structure. In this section, we use synthetic stellar populations to build intuition about the age distributions of stars falling in each of these CMD regions, and to serve as context for the reader. 

\newcolumntype{t}{!{\extracolsep{10pt}}l!{\extracolsep{0pt}}}
\newcolumntype{p}{!{\extracolsep{10pt}}r!{\extracolsep{0pt}}}
\newcolumntype{q}{!{\extracolsep{28pt}}l!{\extracolsep{0pt}}}
\newcolumntype{s}{!{\extracolsep{28pt}}r!{\extracolsep{0pt}}}

\begin{deluxetable}{ttp}[!t]
\tablecaption{\textnormal{MATCH Parameters for Synthetic Catalog}\label{tab:match}}
\tablecolumns{3}
\setlength{\extrarowheight}{2pt}
\tabletypesize{\small}
\tablehead{%
\colhead{Parameter} &
\colhead{Description} &
\colhead{Value} \vspace{-5mm}\\
}
\startdata
\texttt{IMF} & IMF Slope & 1.3 \\ 
\texttt{(m-M)$_0$} & Dist.\ modulus & 24.67 \\
\texttt{diskAv} & Differential Extinction & 0.3,1.0,0.3,0.5 \\
\texttt{Zspread} & Metallicity spread & 0.3 \\
\texttt{BF} & Binary fraction & 0.3 \\
\texttt{dmag\_min} & Min.\ out$-$in mag & $-$0.75 \\
\texttt{SFR} & Constant SFR in $M_{\odot}\,{\rm yr}^{-1}$ & 1.0 \\
\texttt{Z} & Metallicity in each time bin & \citetalias{beasley2015} AMR \\
\enddata
\tablecomments{Parameter values used with the MATCH \texttt{fake} utility to generate the synthetic stellar catalog.}
\vspace{-20pt}
\end{deluxetable}

\begin{deluxetable}{qsss}[t]
\tablecaption{\textnormal{Population Ages}\label{tab:ages}}
\tablecolumns{4}
\setlength{\extrarowheight}{2pt}
\tabletypesize{\small}
\tablehead{%
\colhead{} &
\colhead{16\%} &
\colhead{50} &
\colhead{84\%} \vspace{-2.5mm}\\
\colhead{Population} & 
\colhead{(Gyr)} & 
\colhead{(Gyr)} & 
\colhead{(Gyr)} \vspace{-5mm}\\ 
}
\startdata
MS & 0.014 & 0.050 & 0.112 \\
HeB & 0.050 & 0.141 & 0.251 \\
AGB & 0.45 & 1.12 & 2.00 \\
RGB & 1.26 & 3.98 & 8.92 \\
\enddata
\tablecomments{Median ages and 16--84\% ranges derived for synthetic stars belonging to the four different population selections. The values listed for the RGB are for the `inner' (higher-density) RGB selection, which is uniform across the disk.}
\end{deluxetable}

We used the \texttt{fake} utility of the MATCH SFH-fitting code \citep{dolphin2016} to generate a catalog of stars similar to those found in M33. We first used the \texttt{makefake} utility to generate a synthetic artificial star catalog, of the form described in \cite{williams2021}, and then used \texttt{fake} to sample this synthetic artificial star catalog for an assumed SFH. We used the updated Padova stellar evolutionary model suite, including modeling of the Thermally-Pulsating AGB phase (TP-AGB) and the ratio of Carbon-rich to Oxygen-rich AGB stars \citep{cioni2006a,cioni2006b,marigo2008,girardi2010}, to sample stars from a constant SFH of 1\,$M_{\odot}$\,yr$^{-1}$\ with $\log_{10}$[Age(yr)]\,=\,6.0--10.1 (in 205 0.02\,dex age bins). We assumed the age--metallicity relation measured by \cite{beasley2015} for M33 star clusters --- [M/H]\,=\,$-$1.0 at early times, to a maximum of $-$0.2 for stars $<$\,1\,Gyr old --- with a 0.3\,dex metallicity spread. We assumed a combined foreground and \textit{in situ} differential extinction model, using the \texttt{diskAv} parameter. A flat differential foreground was assumed, up to $A_{V{\rm\ FG,max}}\,{=}\,0.3$, as well as a log-normal distribution affecting all disk stars, with a mean of $\mu(A_{V})_{\rm disk}\,{=}\,0.3$\ and a standard deviation of 0.5\,mag. This dust model is a good approximation of the distribution of extinction values found across the PHATTER footprint by \cite{lazzarini2022}.\footnote{Future work by the PHATTER survey will construct a true map of dust extinction in M33, comparable to the \cite{dalcanton2015} analysis of M31.} We assumed a stellar-to-gas scale height ratio of 1.5 for this distribution. Table \ref{tab:match} gives a comprehensive list of the MATCH parameters used, allowing the reader to recreate these plots if desired.

\begin{figure*}[!ht]
\centering
\leavevmode
\includegraphics[width={\linewidth}]{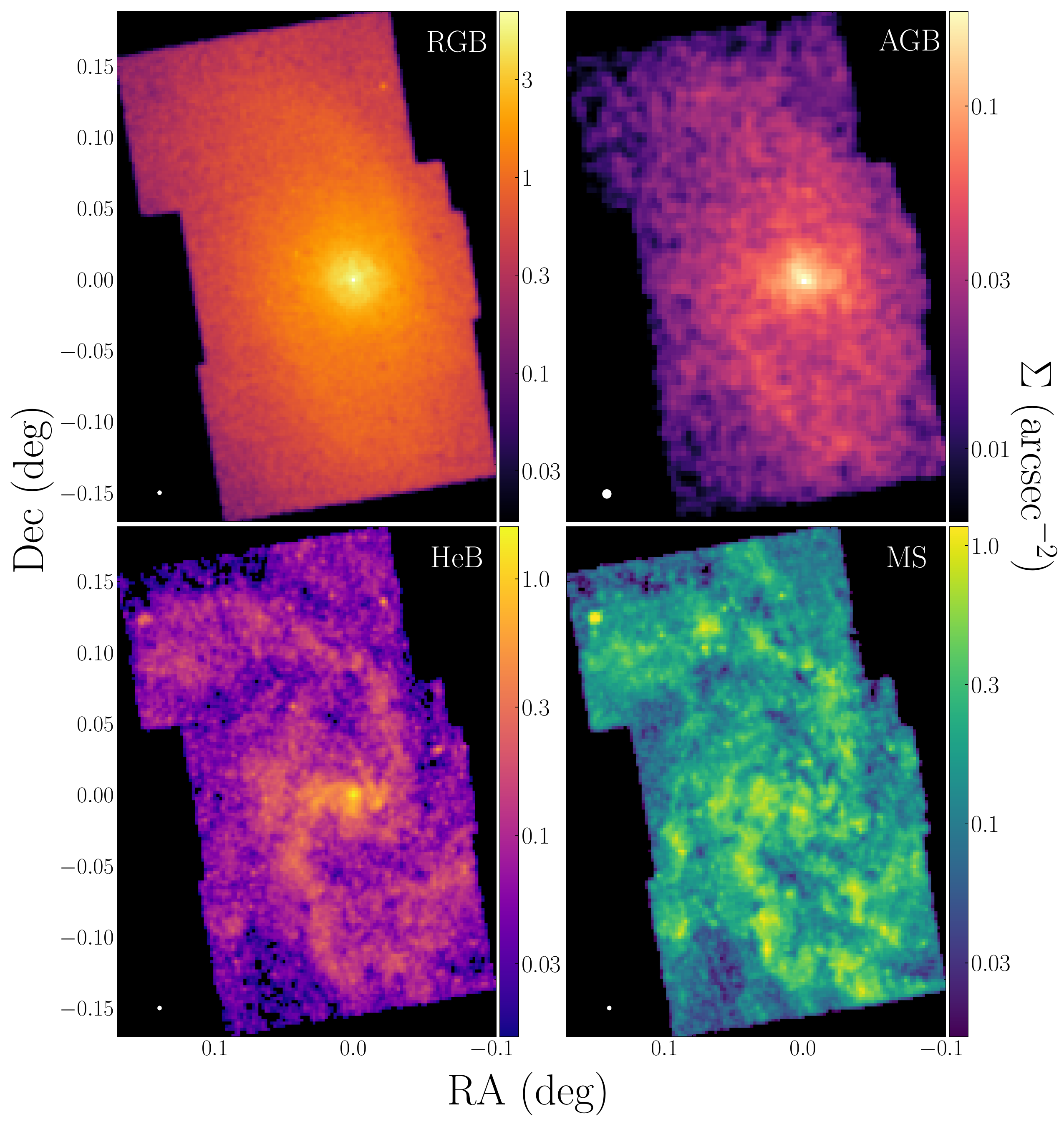}
\caption{Maps of stellar density for each of the four stellar populations defined in S\,\ref{sec:fakepop}: RGB (top left), AGB (top right), HeB (bottom left), and MS (bottom right). All maps are smoothed with a Gaussian kernel (0.65$\sigma$\ for RGB, HeB, and MS; 0.8$\sigma$\ for AGB). A white circle corresponding to the FWHM of the smoothing kernel used is shown in the bottom left of each map. All maps are displayed using a logarithmic scaling in units of stars per arcsec$^2$. The RGB, HeB, and MS maps are shown at a resolution of 7\farcs8 per pixel, while the AGB map is shown at lower resolution of 12\farcs7 per pixel.}
\label{fig:maps}
\end{figure*}

Figure \ref{fig:fakestars} (left and center panels) shows optical and infrared Hess CMDs (equivalent to Figure \ref{fig:selection}) of the synthetic stellar catalog described above. The CMDs are color-mapped to the weighted average age of stars within each color--magnitude bin. The optical CMD is scaled to better display the age gradient in the young populations (i.e., MS and HeB), while the infrared CMD is scaled to better display the age gradient in the old populations (i.e., AGB and RGB). Figure \ref{fig:fakestars} also shows the full age distributions\footnote{Estimated using SciPy's \texttt{stats.gaussian\_kde} function.} of synthetic stars selected as MS, HeB, AGB, and RGB, following the criteria given in Table \ref{tab:select}. We report the median age and 16--84\% range for each population in Table \ref{tab:ages}. 

As expected, the MS stars are youngest, followed by the HeB, AGB, and RGB, respectively. The median ages for each population are relatively evenly spaced in age, in $\sim$0.5\,dex intervals. There is clearly some contamination in the RGB selection from younger HeB and AGB stars, noticeable as a young `tail' of the distribution, but it is a small effect --- $<$\,5\% in this case. We also acknowledge that the accuracy of the age estimates we obtain here will, of course, be impacted by the true shape of M33's SFH. M33's recent SFH appears to be close to constant over the past $\sim$600\,Myr \citep[e.g.,][]{lazzarini2022}; however its global ancient SFH is still uncertain. In the HST fields of \cite{williams2009}, M33's SFH does appear to be structured over time and vary spatially. We consider the impact of two additional limiting SFH cases (described in full in \textsc{Appendix} \ref{sec:alt-sfhs}) on these age estimates, and find our selections to be robust. The age distributions of stars in our MS and HeB selections are insensitive to changes to the ancient SFH, while the ages of stars in the AGB selection fluctuate by $\sim$40\%. The most significant changes are exhibited by stars in the RGB selection, as expected. Regardless, our RGB selection is reliably tracing ancient stellar populations with typical ages $\gg$1\,Gyr. 

We experimented with selecting the younger stellar populations in the NIR as well, to explore the effects of dust, which could be substantial for these younger, more embedded populations. However, we found very little difference ($<$5\%) in the total number of selected stars or the visual morphology obtained for these two selections, suggesting that extinction does not impact our morphological inferences.

\section{Results}
\label{sec:results}

In this section, we present the results of our analysis of M33's structure. In \S\,\ref{sec:density} we present maps of stellar density for each of the four populations selected in \S\,\ref{sec:select}, followed by a detailed analysis of the detection and characterization of M33's spiral structure and a first-ever confirmation of a central bar in \S\,\ref{sec:spiral}. Finally, in \S\,\ref{sec:decomp} we fit elliptical isophotes to the RGB density distribution (\S\,\ref{sec:isophote}), extending our coverage with existing near-infrared broadband imaging, followed by a multi-component Markov chain Monte Carlo (MCMC) decomposition of the resulting radial density profile (\S\,\ref{sec:decomp-fit}).

\subsection{Population Density Maps}
\label{sec:density}

In Figure \ref{fig:maps} we present maps of stellar density for each of the four populations described in \S\,\ref{sec:select}: RGB, AGB, HeB, and MS. The RGB stars are the most numerous and are subject to severe crowding in M33's inner regions. Therefore, to probe the density structure of M33 independent of completeness effects, we consider two selection schemes for RGB stars: an `outer' selection for stars with projected circular radii $>$0\fdg02 from M33, and an `inner', shallower selection for stars with projected circular radii $\leqslant$0\fdg02 from M33 --- corresponding to the region of highest stellar density in \cite{williams2021}. We give the criteria for these two RGB selections in Table \ref{tab:select}. To effectively blend these two different selections, we select stars using each selection in a thin, 0\farcs71 circular annulus at a 0\fdg02 radius from M33, and take the ratio of the number of stars obtained with each selection scheme. When calculating the RGB density map (and throughout the rest of the paper), we then weight each `inner'-selected RGB star by this scale factor of 2.45. The RGB, HeB, and MS maps are shown at a square pixel scale of 7\farcs8 per pixel, or $\sim$32.5\,pc per pixel in projected distance units. The AGB map is shown at a lower resolution compared to the other maps: 12\farcs7 per pixel ($\sim$53\,pc per pixel projected). 

\begin{figure}[t]
\centering
\leavevmode
\includegraphics[width={\linewidth}]{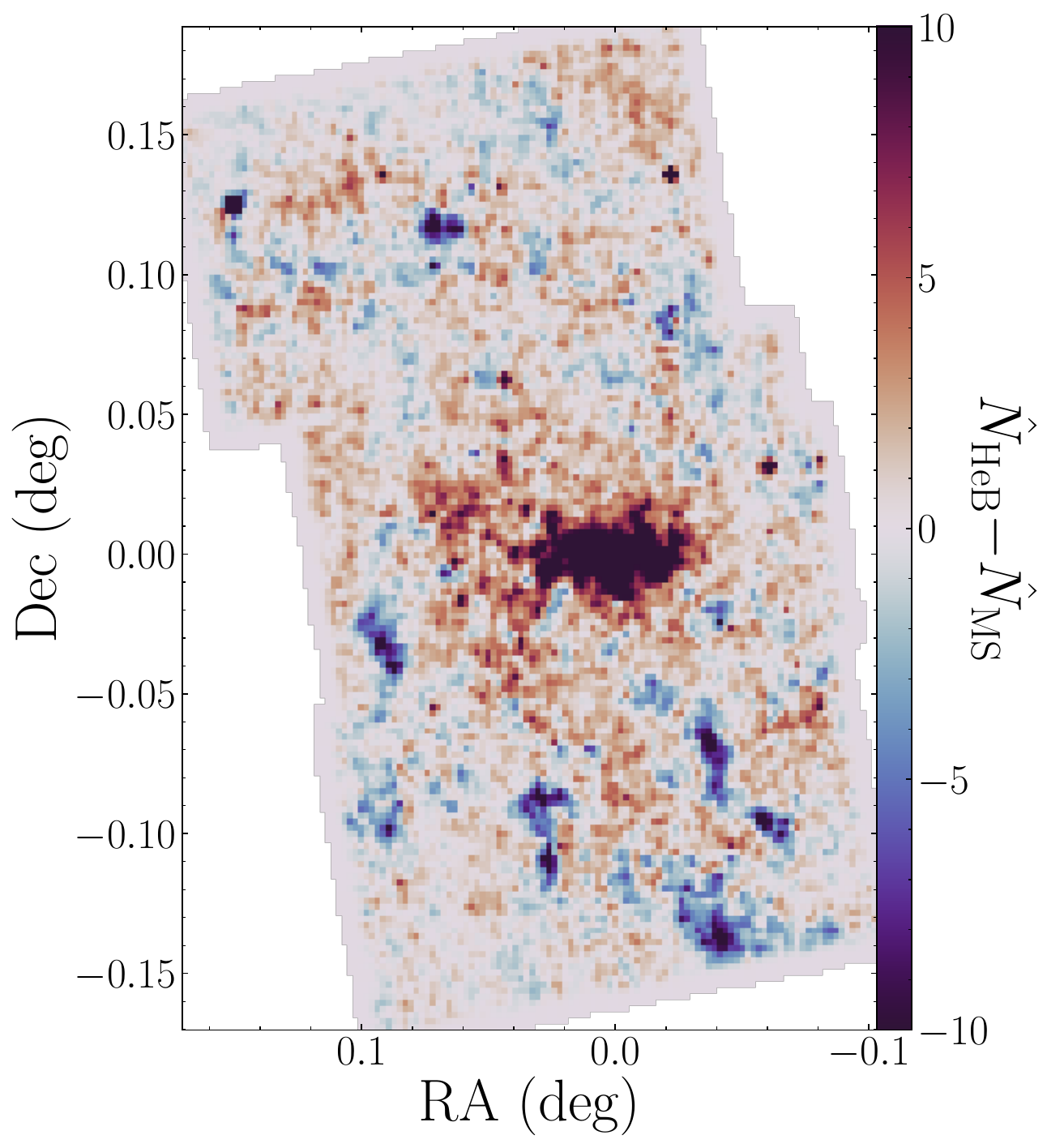}
\caption{Difference map of HeB and MS stars, scaled to the same expected total stellar mass in each population. The map is shown at the same 7\farcs8 resolution as Figure \ref{fig:maps}.}
\label{fig:heb-ms}
\end{figure}

Surprisingly, the MS map is the only one to exhibit the flocculent spiral structure typically attributed to M33. The HeBs show similar structures, but two spiral arms stand out more already in this map (as noted by \citealt{humphreys&sandage1980} and \citealt{regan&vogel1994}). They also show a bar-like structure, similar to the structure observed in the star formation maps of \cite{lazzarini2022} at intermediate ages. Figure \ref{fig:heb-ms} shows the difference between the HeB and MS maps, normalized to the expected stellar mass formed in each selection region,  defined using the synthetic constant SFH CMDs shown in Figure \ref{fig:fakestars}. While these are different age populations, and therefore should not be expected to be distributed identically, the bar and southern primary spiral arm are clearly more prominent in the intermediate age HeB stars, relative to the younger upper MS populations.

The AGB stars exhibit a purely barred, two-arm structure, without any visible flocculence and a large contribution from a smooth disk. The RGB stars largely mirror the structure of the AGB, though the bar structure is less visible against the dominant smooth disk, with the two spiral arms representing a modest density enhancement. The following section is dedicated to a more quantitative characterization of the visible bar and spiral arms.

\subsection{Detecting the Bar and Spiral Arms}
\label{sec:spiral}

\begin{figure*}[t]
\centering
\leavevmode
\includegraphics[width={0.8\linewidth}]{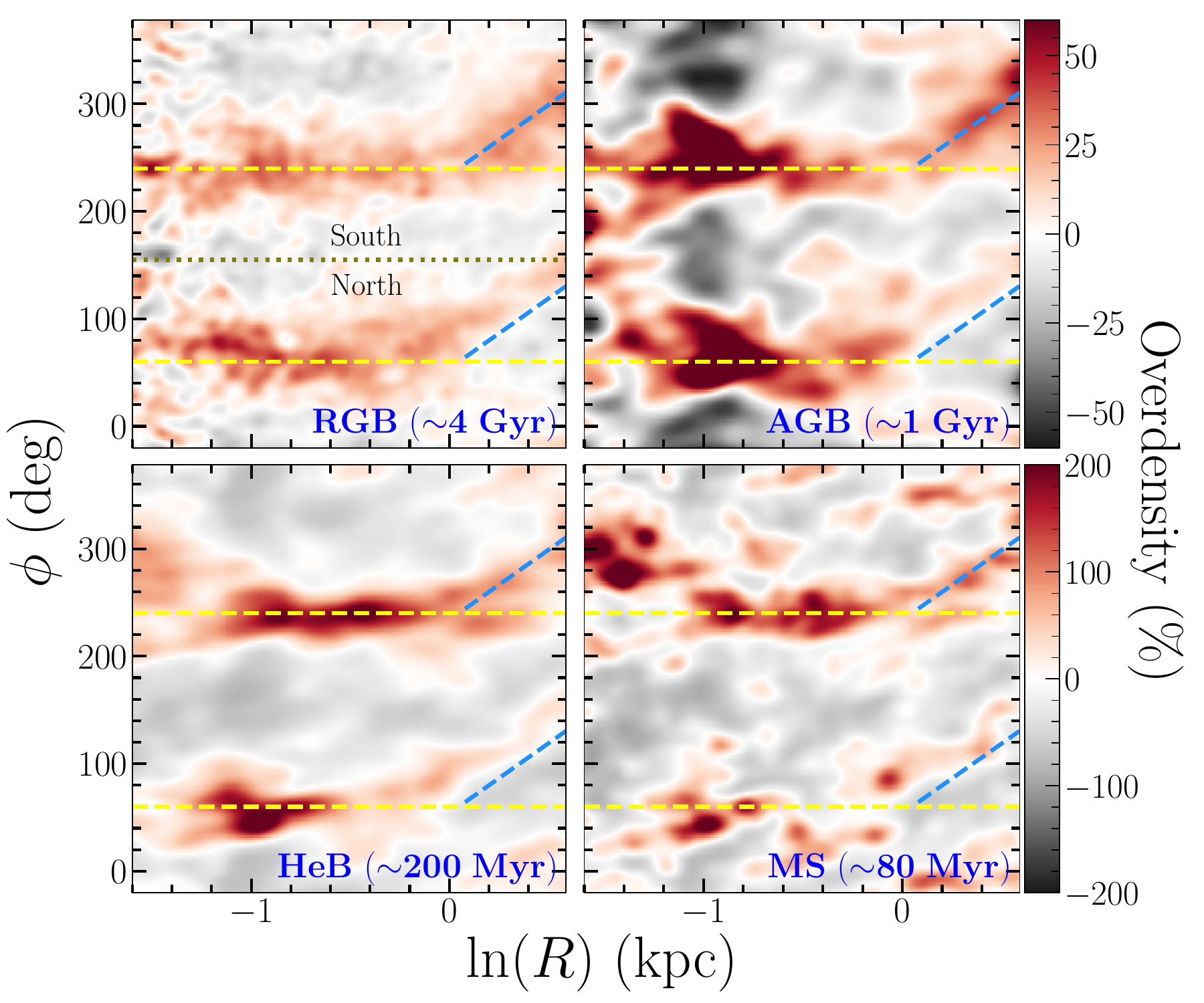}
\caption{Elliptical phase angle plotted against natural logarithm of radius for stars belonging to each of the four sub-populations. The average age determined from the constant SFH synthetic CMDs for each population is given in blue (see Table \ref{tab:ages}). Each row is displayed with the same color scaling, corresponding to the two colorbars to the right. The range of $\phi$\ has been extended by 50\textdegree\ at the top and bottom of the $y$-axis, to improve the continuity of visible structures. A fixed ellipse with ellipticity, ${\epsilon}\,{=}\,0.43$\ (corresponding to an inclination of 55\textdegree), and position angle, ${\theta}\,{=}$\,201\fdg1, was assumed \citep[e.g.,][]{koch2018}. $\phi$\ is defined starting from M33's western minor axis, and is measured clockwise in RA (counterclockwise visually). Each panel has been rescaled as a percentage overdensity relative to the per-radial-bin mean. Dashed lines representing the bar structure are shown in yellow, while the blue dashed lines denote the two dominant spiral arms. The blue lines are 180\textdegree\ apart, measured from the southeastern spiral arm at ${\phi}\,{\sim}$\,240\textdegree, though the northwestern arm is offset from this assumption of perfect symmetry.}
\label{fig:spiral}
\end{figure*}

\begin{figure}[!t]
    \centering
    \includegraphics[width={\linewidth}]{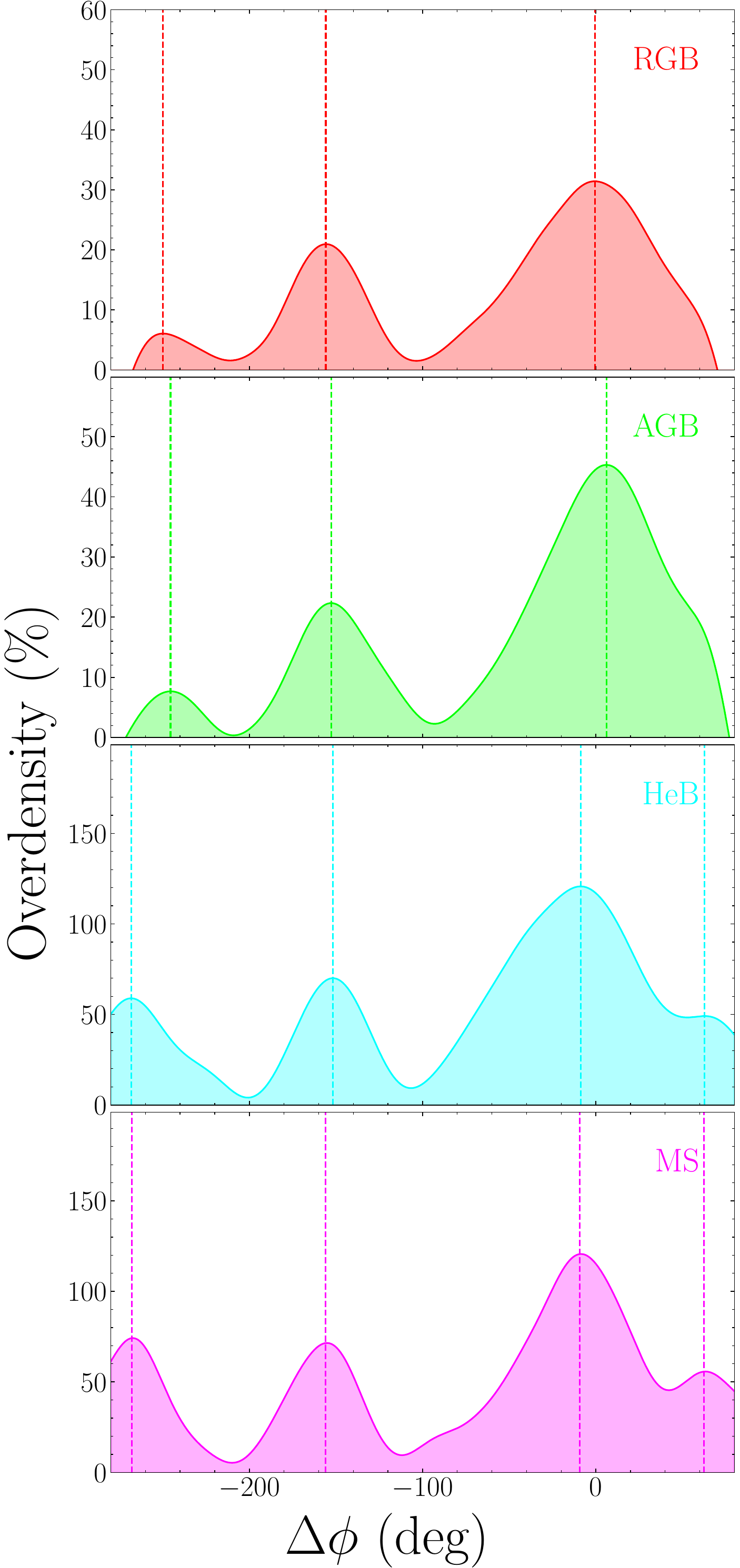}
    \caption{Distributions of the positive component of the phase angle offset from M33's southeastern spiral arm (relative to the upper blue line in Figure \ref{fig:spiral}), for stars with $\ln(R)\,{>}\,0$\ (1\,kpc), in each of the four sub-populations. The  centers of identified peaks in each panel are denoted by dashed lines in the corresponding color. As in Figure \ref{fig:spiral}, the smooth radial density component has been removed.}
    \label{fig:sp-offset}
\end{figure}

To characterize the structures observed in the density maps, such as the visible spiral arms and possible bar, we convert the ($\Delta\alpha$, $\Delta\delta$) positions of stars to coordinates of elliptical radius and phase angle. The relative phase angle, $\phi$, and radius, $R$, of a point along a rotated ellipse are given by:
\begin{equation}
\begin{aligned}
\phi ={} & \arctan^{-1}\bigg(\frac{1}{1-\epsilon}\,\frac{x\,\sin\theta - y\,\cos\theta}{x\,\cos\theta + y\,\sin\theta}\bigg) \\
R ={} & \Big((x\,\cos\theta + y\,\sin\theta)^2 + \frac{1}{(1-\epsilon)^2}(x\,\sin\theta - y\,\cos\theta)^2\Big)^{1/2}, \\
\end{aligned}
\end{equation}
where ($x$, $y$)\,$\equiv$\,($\Delta\alpha$, $\Delta\delta$), $\epsilon$\ is the ellipticity and $\theta$\ is the position angle of the ellipse. This is particularly useful in the analysis of structures that are anisotropic with the phase of the ellipse. For example, logarithmic spiral arms will appear as a diagonal line in $\ln(R){-}\phi$\ space, while a bar will appear as a flat line, as it has an approximately fixed phase angle across a range of radii. Additionally, components that are approximately isotropic with $\phi$, such as the disk or a central spheroid, can be easily subtracted as the mean density in each radial bin without fitting them analytically. In calculating $\phi$\ and $R$, we take $\theta\,{=}$\,201\fdg1 (relative to due south, rotating clockwise with RA, or counterclockwise on the sky) and $\epsilon\,{=}\,0.43$, corresponding to an inclination angle of $i\,{=}\,55$\textdegree, from \cite{koch2018}.

Figure \ref{fig:spiral} shows a kernel density estimate (KDE) of sources in $\ln(R){-}\phi$\ space, evaluated on a grid with a resolution of 0.01 in $\ln(R)$\ and 2\textdegree\ in $\phi$. In each panel, we subtracted the average density at each radius, across all phase angles, to better showcase density enhancements. For the RGB and AGB, this subtracted component largely represents the disk, which is dominant in Figure \ref{fig:maps}. The HeB and MS stars are more structured, resulting in larger density enhancements relative to the mean. The RGB, AGB, and HeB all show very clearly a bar structure, extending out to $\sim$1--1.5\,kpc, as well as a pair of spiral arms. The MS stars exhibit a much weaker bar structure, and a greater number of spiral features, as seen in Figure \ref{fig:maps}. 

\begin{figure*}[t]
\centering
\leavevmode
\includegraphics[width={0.7\linewidth}]{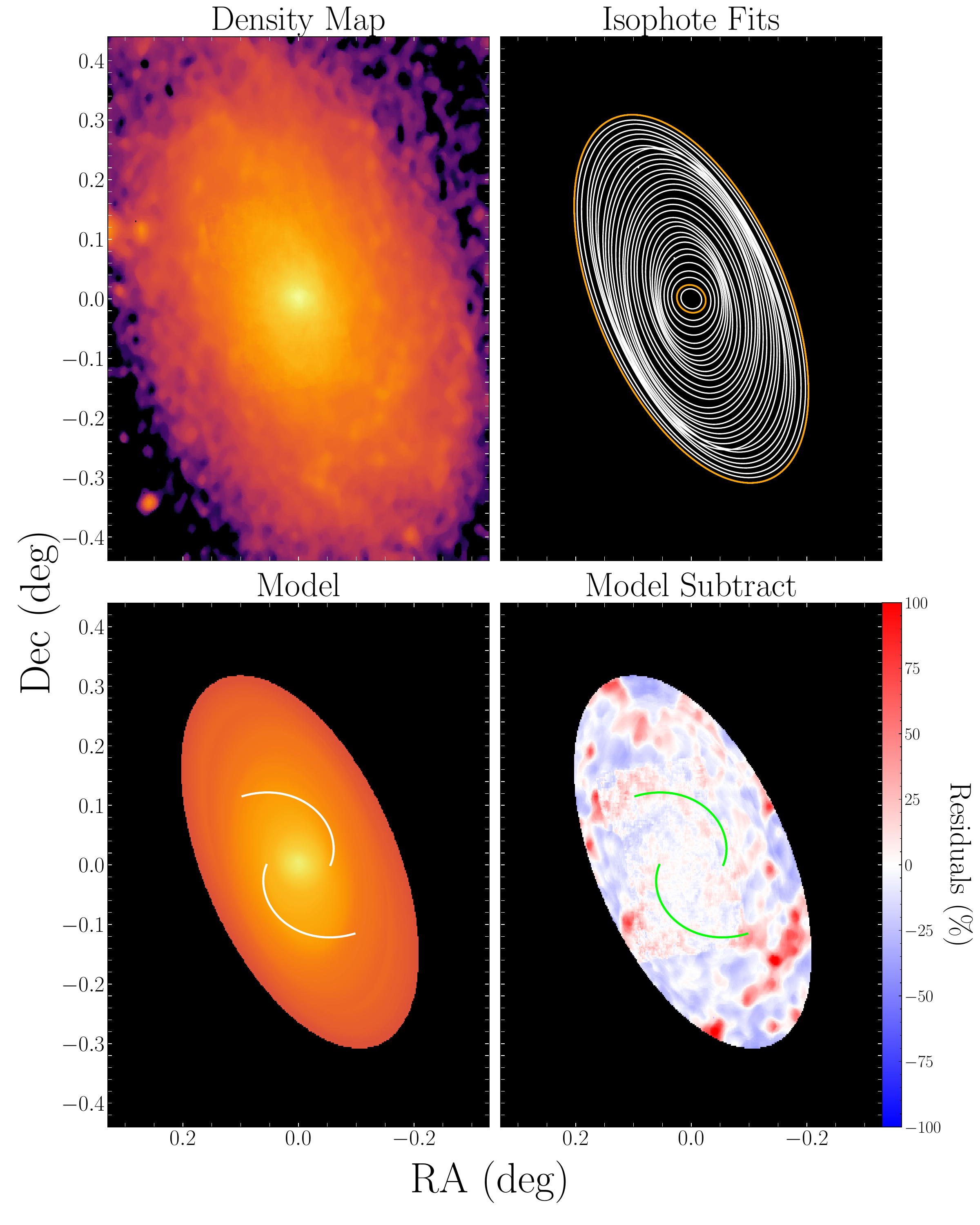}
\caption{Results of the elliptical isophote fitting procedure. The top left shows the smoothed RGB density map, combined with the scaled 3.6\,\um\ Spitzer image. The top right shows the best fit elliptical isophotes at each radius. The inner bar and outer disk isophotes are shown in orange, from which we estimate the bar strength (see \S\,\ref{sec:decomp}). The bottom left shows the best-fit smooth 2D model, and the bottom right shows the model data$-$model residuals, expressed as a percentage relative to the model. White/green curves in the lower two panels correspond to logarithmic spiral functions with pitch angle equivalent to the two dominant spiral features highlighted in Figure \ref{fig:spiral}. As can be seen, large-scale density residuals are typically within 25--30\%, with larger residuals almost exclusively residing in regions dominated by spiral arms and/or active star-forming clumps.} 
\label{fig:elliptical}
\end{figure*}

The two primary spiral features visible in all four panels of Figure \ref{fig:spiral} are $<$180\textdegree\ apart in phase angle, and thus are asymmetric compared to the expectation of a perfect logarithmic spiral, though they do appear to exhibit the same slope. As the radius of a logarithmic spiral takes the form 
\begin{equation}
    R(\phi) = R_0\,e^{\tan(\theta_{p})\phi},
\end{equation}
where $\theta_{p}$\ is the spiral pitch angle, the pitch angle of M33's spiral arms is given by the arctangent of the inverse slope of the linear features in Figure \ref{fig:spiral}. For an estimated slope of 2.27, this gives a pitch angle of $\theta_{p}\,{=}$\,23\fdg78, with an inner radius of $R_0\,{=}\,0.90$\,kpc. Though these are just the innermost spiral features in M33, this is typical of the spiral pitch angles of other late-type galaxies \citep[e.g.,][]{diaz-garcia2019}. In Figure \ref{fig:sp-offset}, we show the distribution of phase angle offsets for each population (with the disk and other symmetric components subtracted), relative to the plane of the logarithmic spiral model for the upper arm in Figure \ref{fig:spiral}, which appears to be the strongest overall. 

As expected from Figure \ref{fig:spiral}, we find two dominant peaks in the RGB and AGB distributions, which are separated by $<$180\textdegree. We find a separation of $\sim$155--160\textdegree\ between these two peaks in the RGB stars, or $\sim$20--25\textdegree\ offset from perfect symmetry. These two features appear with a similar separation in all three other populations. This offset is comparable to the phase asymmetry found in the spiral arms of M51 (20\textdegree), which is thought to be caused by an additional multi-arm spiral density wave triggered by its interaction with M51b \citep{henry2003}. Furthermore, the two primary peaks are asymmetric in strength, as visually apparent in Figure \ref{fig:spiral}, which also supports recent interaction as an important driver of M33's structure. The remaining phase decomposition shows more complex structure, with multiple overlapping modes visible for the younger HeB and MS stars. 

\subsection{Structural Decomposition}
\label{sec:decomp}

In this section, we conduct elliptical isophote fitting of the distribution of M33's RGB stars, which are presumably the best tracers of stellar mass. Using the best-fit model, we conduct an MCMC-based structural decomposition of the resulting radial profile, to explore the components of M33's structure in greater detail. 

\subsubsection{Isophote Fitting}
\label{sec:isophote}

To fit elliptical isophotes to our RGB density map of M33 we used the \texttt{isophote}\footnote{An implementation of the algorithm described in \cite{jedrzejewski1987}, also used in \texttt{IRAF}'s \texttt{Ellipse} function.} utility of \texttt{photutils}, a Python package created by the Astropy Project for photometric analysis \citep{larry_bradley_2020_4044744}. The PHATTER HST imaging only covers the inner $\sim$2.5\,kpc of M33's disk, and thus we extend the radial range of our analysis to better probe M33's outer disk component. We do so by combining the density map of resolved RGB stars with the archival \textit{Spitzer} IRAC 3.6\,\um\ mosaic of M33 produced by the Local Volume Legacy Survey \cite{dale2009}. We scaled the 3.6\,\um\ image by the ratio of median RGB density (in arcsec$^{-2}$) to median surface brightness (in MJy\,sr$^{-1}$) in the PHATTER footprint --- 0.66 RGB\,arcsec$^{-2}$/MJy\,sr$^{-1}$. We then masked bright foreground stars and median-smoothed the 3.6\,\um image to remove fine structure from star forming regions, including bright red supergiants \citep[e.g.,][]{regan&vogel1994} and emission from warm dust grains that contaminate near-infrared broadband imaging. In Figure \ref{fig:elliptical} (top left) we show the combined smoothed RGB count/3.6\,\um\ map used for fitting. The resulting map shows no discontinuity associated with the change from RGB number counts to the Spitzer image. 

We fit the combined density image with elliptical isophotes in 150 linearly-stepped bins in semi-major axis length with size 0\farcm126, out to a maximum semi-major axis length of 18\farcm9 ($\sim$0.03--4.72\,kpc). For each isophote, we allowed the the center position, ellipticity, and position angle to vary. In Figure \ref{fig:elliptical} we show the fitted ellipses at each radius (top right), the reconstructed model image (bottom left), and the model-subtracted residuals (bottom right). The model reproduces the smooth distribution of M33 quite well, with bulk residuals $\lesssim$30\% (except for a number of higher-residual features visibly associated with spiral structure). The residuals within the PHATTER footprint are $<$\,20\%. 

\begin{figure*}[t]
\centering
\leavevmode
\includegraphics[width={\linewidth}]{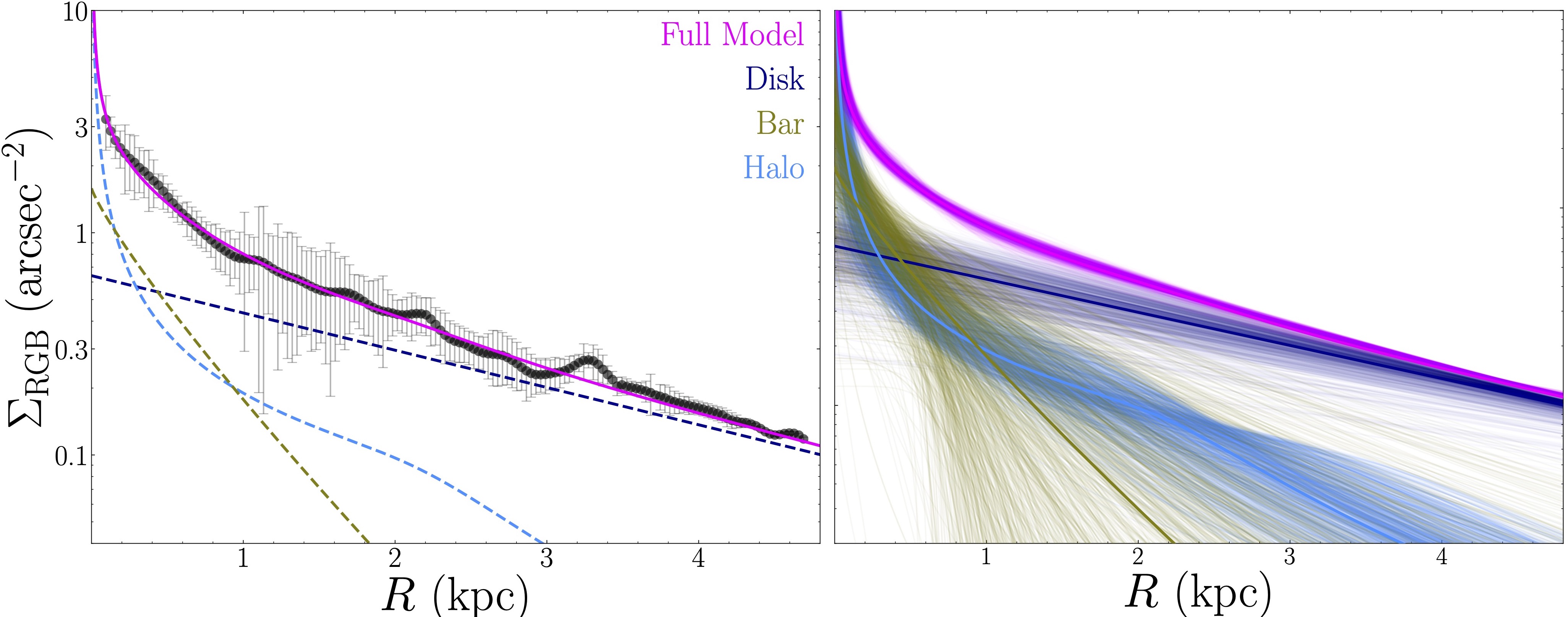}
\caption{\textit{Left}: RGB density radial profile, measured using the elliptical apertures obtained from the isophote fitting. Points at $R\,{>}\,2.6$\,kpc are obtained from the Spitzer 3.6\,\micron\ imaging. The best-fit model from our MCMC fitting procedure is shown overtop in pink, with the three individual components shown as dashed curves. \textit{Right}: The best-fit total model and each component are shown as colored solid curves, while 1000 randomly-sampled realizations for each from the MCMC run are shown as transparent curves of the same color.}
\label{fig:profile}
\end{figure*}

Applying the fitted isophotes to the combined image in each radial bin, we measure the RGB stellar density within each elliptical annulus and plot the extracted radial density profile for M33 in Figure \ref{fig:profile} (left). The profile was modestly smoothed with a third-order Savitzky--Golay filter. Error bars represent the quadrature-sum of Poisson uncertainty on each count, and the standard deviation in density of 300 bootstraps of the elliptical parameters for each isophote. For the bootstrapped density estimates, each parameter was assumed to be normally distributed, with the width taken as the 1$\sigma$\ (68\%) uncertainty on each parameter output by \texttt{photutils}.

The profile has a number of recognizable features, including a sharp inner slope change around $\sim$1\,kpc, an exponential-dominated region, and a central upturn. We decompose the profile into its constituent morphological components in the following section. 

\subsubsection{Profile Decomposition}
\label{sec:decomp-fit}

Taking the radial profile derived from fitting elliptical isophotes, and the insight gained from analysis of M33 in the phase diagrams presented in Figure \ref{fig:spiral}, we conducted a decomposition with three assumed morphological components: (1) an exponential disk, (2) a S\'{e}rsic component representing the bar (see \S\,\ref{sec:spiral}), and (3) a spheroidal broken power law component that we will show is consistent with the properties of an accreted `halo' at both small and large radii. 

\newcolumntype{m}{!{\extracolsep{-25pt}}l!{\extracolsep{0pt}}}
\begin{table*}
\begin{tabular*}{0.5\textwidth}{l m}
    & \vbox{\hspace{85pt}\textbf{Disk} \begin{equation}
        \Sigma_{\rm disk}(r) = \Sigma_{\rm e,disk}\,\exp\Bigg\{{-}1.678\,\bigg[\frac{r}{R_{\rm e,disk}} - 1\bigg]\Bigg\}
        \label{eq:disk}
        \end{equation}} \\
    & \vbox{\hspace{85pt}\textbf{Bar} \begin{equation}
        \Sigma_{\rm bar}(r) = \Sigma_{\rm e,bar}\,\exp\Bigg\{{-}b_n\,\bigg[\bigg(\frac{r}{R_{\rm e,bar}}\bigg)^{1/n_{\rm bar}} - 1\bigg]\Bigg\}
    \label{eq:bar}
    \end{equation}} \\
    & \vbox{\hspace{85pt}\textbf{Halo} \begin{equation}
        \Sigma_{\rm halo}(r) = \Sigma_{\rm 0,halo}\,\bigg(\frac{r}{R_{\rm break}}\bigg)^{\alpha_1}\,\Bigg\{0.5\,\Bigg[1 + \bigg(\frac{r}{R_{\rm break}}\bigg)^{1/\Delta}\Bigg]\Bigg\}^{(\alpha_1{-}\alpha_2)\Delta}
    \label{eq:halo}
    \end{equation}} \vspace{12pt}\\ 
    & \vbox{\hspace{85pt}\textbf{Total} \begin{equation}
        \Sigma(r) = \Sigma_{\rm disk}(r) + \Sigma_{\rm bar}(r) + \Sigma_{\rm halo}(r)
    \label{eq:total}
    \end{equation}} \\
\end{tabular*}
\tablecomments{Functional forms for each of the three components assumed in our radial profile decomposition. The Disk and Bar components are expressed as S\'{e}rsic functions ($n\,{=}\,1$\ for the disk), while the halo component is a broken power law (taken from \texttt{astropy.modeling}'s \texttt{SmoothlyBrokenPowerLaw1D} function). }
\end{table*}

\newcolumntype{y}{!{\extracolsep{120pt}}l!{\extracolsep{0pt}}}
\newcolumntype{z}{!{\extracolsep{60pt}}r!{\extracolsep{0pt}}}

\begin{deluxetable}{lz}[t]
\tablecaption{\textnormal{Density Profile Fit Results}\label{tab:fit}}
\tablecolumns{2}
\setlength{\extrarowheight}{4pt}
\tablewidth{\linewidth}
\tabletypesize{\small}
\tablehead{%
\colhead{Parameter} & 
\colhead{Best Fit}}
\startdata
\multicolumn{2}{l}{\textbf{Disk} (Exponential)}\vspace{2pt}\\
\hline
$I_{\rm e,disk}$ (arcsec$^{-2}$) & 0.12$_{{-}0.03}^{{+}0.03}$ \\
$R_{\rm e,disk}$ (kpc) & 4.34$_{{-}0.37}^{{+}0.61}$\vspace{3pt}\\
\hline
\multicolumn{2}{l}{\textbf{Bar} (S\'{e}rsic)}\vspace{2pt}\\
\hline
$I_{\rm e,bar}$ (arcsec$^{-2}$) & 0.21$_{{-}0.08}^{{+}0.19}$ \\
$R_{\rm e,bar}$ (kpc) & 0.90$_{{-}0.42}^{{+}0.71}$ \\
$n_{\rm bar}$ & 1.16$_{{-}0.64}^{{+}0.54}$\vspace{3pt}\\
\hline
\multicolumn{2}{l}{\textbf{Halo} (Broken Power Law)}\vspace{2pt}\\
\hline
$I_{\rm 0,halo}$ (arcsec$^{-2}$) & 0.08$_{{-}0.03}^{{+}0.05}$ \\
$\alpha_1$ & $-$0.90$_{{-}0.18}^{{+}0.17}$ \\
$\alpha_2$ & $-$3.07$_{{-}0.34}^{{+}0.41}$ \\
$R_{\rm break}$ (kpc) & 2.29$_{{-}0.58}^{{+}0.91}$ \\
$\Delta$ & 0.1\vspace{3pt}\\
\enddata
\end{deluxetable}

It is obvious, particularly given our relatively small radial coverage (only extending out to $\sim$7\,kpc with the Spitzer imaging), that our choice of a single broken power-law component to model M33's halo is degenerate with other combinations of morphological components, including a pure exponential at larger radii and a spherical S\'{e}rsic bulge to explain the central upturn at smaller radii. However, in a large Keck DEIMOS spectroscopic campaign targeting 1667 identified bright RGB stars in M33, \cite{gilbert2022} detected a substantial high-velocity dispersion, non-rotating stellar component that persists over a broad range of radii --- similar to the properties of a stellar halo. While the term `stellar halo' is most often used to describe the diffuse outskirts of galaxies that are dominated by stars accreted from other, smaller galaxies, most models predict that accreted stellar halos can be described as power law distributions even to very small radii, well within the main body of the hsot galaxy \citep[e.g.,][]{bullock&johnston2005,monachesi2019}, and this has been born out in observations of the MW and nearby galaxies \citep[e.g.,][]{deason2011,harmsen2017}. We therefore model this high velocity dispersion component in M33 as a power law, and will refer to it as a `halo' for the remainder of the paper for simplicity.\footnote{As the inhomogeneous spatial coverage of \cite{gilbert2022} precludes measurement of this halo component's axis ratio, we assume this component to be spherical.} We stress that the inclusion of a third morphological component to describe the density profile is entirely motivated by the \cite{gilbert2022} discovery of a hot component, and not by ``goodness of fit'' concerns. See \textsc{Appendix} \ref{sec:singleplaw} for additional discussion to this point. 

The dynamics of this halo component are consistent for both the innermost and outermost stars, suggesting that it is a single population. Moreover, \cite{gilbert2022} observe a relatively steep decline in the fraction of stars halo component with radius, with a fraction of $>$30\% of the total stellar mass at the innermost radii probed by the sample. These constraints motivate our choice of a broken power law, as any halo component must simultaneously exhibit both a high mass fraction in the inner disk and a low mass fraction in the outer disk. The only model that accomplishes this is a power-law that breaks at relatively small radii. There is precedent for broken power-law stellar halos in our own Milky Way \citep[e.g.,][]{deason2011}, and in hydrodynamical simulations \citep[e.g.,][]{bullock&johnston2005,amorisco2017,monachesi2019}. Given that a power-law and a S\'{e}rsic model both result in a central cusp, we assume that the inner power-law can therefore also be used to fit the central upturn in M33's density profile, rather than assuming a separate, very small bulge component. We explore the validity of our choice of a broken vs.\ single power-law model in \S\,\ref{sec:halo} \&\ \textsc{Appendix} \ref{sec:singleplaw}, and note that the bar and halo components do not appear to be strongly degenerate (see Figure \ref{fig:corner}). 

Equations \ref{eq:disk}--\ref{eq:halo} give the functional forms for each component, while Equation \ref{eq:total} gives the full source density model as a function of radius. To estimate the probability of the observed density profile given our three-component model we adopt the generalized likelihood function \citep[following][]{hoggbovylang2010}:
\begin{equation}
    \ln \mathcal{L} = -\frac{1}{2} \sum_{i=1}^n \Bigg\{\frac{\big[\Sigma_{i,{\rm RGB}} - \Sigma_i(r)\big]^2}{\sigma_i^2} + \ln(\sigma_i^2)\Bigg\}, 
    \label{eq:2}
\end{equation}
$\sigma_i$\ represents the uncertainty on each point in the density profile, including both the Poisson uncertainty on the count and a bootstrapped uncertainty on the elliptical parameters for each isophote (as explained in \S\,\ref{sec:isophote}). We then compute the posterior probability distributions of the model parameters given the data using the MCMC sampler \texttt{emcee} \citep{foreman-mackey2013}. The full corner plot for nine-parameter model is shown in \textsc{Appendix} \ref{sec:singleplaw}. Table \ref{tab:fit} gives the most likely values for each parameter, as well as 16$^{\rm th}$\ and 84$^{\rm th}$\ percentile uncertainties. 

\begin{figure}[t]
\centering
\leavevmode
\includegraphics[width={\linewidth}]{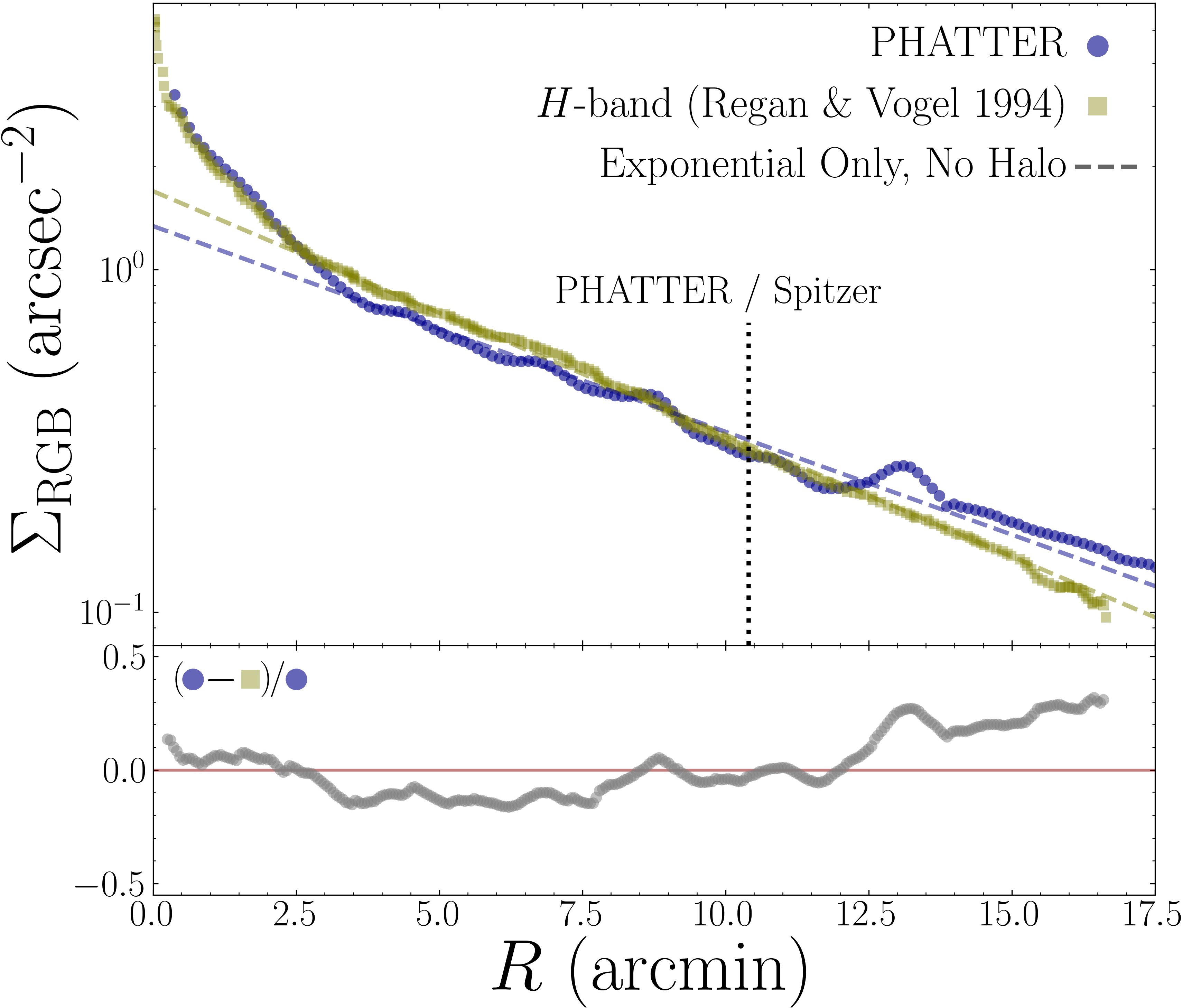}
\caption{A comparison of the density profile calculated in this work using RGB star counts (blue), and the $H$-band surface brightness profile calculated by \cite{regan&vogel1994} (green). The approximate radius at which the Spitzer 3.6\,\micron\ data is used is denoted by a black dashed line. We also show the fractional difference, relative to the RGB density profile, between the two profiles as a function of radius (bottom panel). The $H$-band profile has been scaled to match the PHATTER profile at 2\farcm5. To facilitate a direct comparison to \cite{regan&vogel1994}, we plot single exponential fits to the disk-dominated ($R\,{>}\,3$\arcmin) portions of each profile (dashed lines).}
\label{fig:profile-comparison}
\end{figure}

We find a low-$n$\ S\'{e}rsic component with an effective radius of 0.9\,kpc, consistent with the scale of the bar feature seen in Figure \ref{fig:spiral}, and a broken power-law halo with a reasonably steep outer slope of $-$3.07. Due, in part, to including this underlying halo component, we recover a somewhat larger exponential scale length for M33's disk than previous studies, with an effective radius of 4.34\,kpc compared to 2.56\,kpc in broadband infrared imaging \citep{regan&vogel1994}. If the halo component is excluded, we recover an exponential disk with an effective radius of 3.03\,kpc --- much more consistent with the profile measured in the near-infrared by \cite{regan&vogel1994}, as shown in Figure \ref{fig:profile-comparison} (though their profile appears steeper at large radii), and even more so with the profiles measured in optical filters. It is worth noting that, especially beyond $\sim$10\arcmin, our profile is visibly shallower than the $H$-band profile of \cite{regan&vogel1994}. We have determined that this is likely due to differences in background subtraction between the $H$-band and Spitzer 3.6\micron imaging, mainly: the LVL Spitzer imaging is substantially deeper, detecting real flux at larger radii that was subtracted from the ground-based $H$-band profile. If we adopt a more aggressive background subtraction ($\sim$twice that measured by \citealt{dale2009}), we recover an identical exponential scale length as measured by \cite{regan&vogel1994}.

We characterize the strength of the best-fit bar using a classical, observationally-motivated definition of bar strength in the literature, $f_{\rm bar}$\ \citep{abraham1999,abraham&merrifield2000}, defined as
\begin{equation}
    f_{\rm bar} = \frac{2}{\pi}\Bigg[{\rm arctan}\bigg(\frac{b}{a}\bigg)_{\rm bar}^{-1/2} - {\rm arctan}\bigg(\frac{b}{a}\bigg)_{\rm bar}^{+1/2}\Bigg], 
\end{equation}
where the intrinsic bar axis ratio, $(b/a)_{\rm bar}$, is calculated from the `twist' between an outer disk-dominated isophote and the outermost isophote dominated by the bar. This metric has been shown to be consistent with other metrics for bar strength in the literature \citep[e.g,][]{garcia-gomez2017}. For M33, we show these two isophotes in the top right panel of Figure \ref{fig:elliptical}, highlighted in orange. We measure $f_{\rm bar}\,=\,0.18$\ for M33, which is consistent with the lower end of classically-barred galaxies \citep[$f_{\rm bar}\,{\sim}\,0.15{-}0.4$;][]{abraham&merrifield2000}. 

\section{M33 in Context}
\label{sec:context}
 
In this section, we discuss our results on M33's structure in the context of the rich existing literature on this well-studied galaxy. We first discuss our implications for M33's secular evolution and recent tidal interaction, in \S\,\ref{sec:global-struct}. Then, in \S\,\ref{sec:halo} we discuss of our modeling of M33's stellar halo in the context of previous studies of it's stellar structure at large scales. 

\subsection{M33's Global Structure: A Barely-Stable Disk Shaped by Tidal Interaction}
\label{sec:global-struct}

The results of \S\,\ref{sec:results} represent a dramatic shift in our understanding of M33's structure: rather than a prototypical low-mass flocculent spiral, M33's dominant stellar morphology is more in line with that of a barred, two-arm grand-design spiral. While the presence of a bar in M33 has been long discussed \citep[e.g.,][]{regan&vogel1994,corbelli&walterbos2007,lazzarini2022}, and indeed it has been argued that its disk \textit{should} form a bar \citep{sellwood2019}, these results are the first quantitative evidence for it. Despite its underlying grand design structure, M33's spiral arms exhibit substantial asymmetry, in both strength and orientation. Two primary questions therefore remain: 
\begin{enumerate}[itemsep=-1pt,topsep=3pt,leftmargin=14pt]
    \item What drives the asymmetry in M33's dominant spiral features? 
    \item Why do M33's old and young stars present such different morphologies, and is this common? 
\end{enumerate}

Asked more generally: how have interaction-driven and secular processes shaped M33's structure? We discuss our new insights into each in \S\,\ref{sec:interaction} and \S\,\ref{sec:potential}, respectively.

\subsubsection{A Disk Shaped by Interaction}
\label{sec:interaction}

\begin{figure*}[t]
\centering
\leavevmode
\includegraphics[width={0.65\linewidth}]{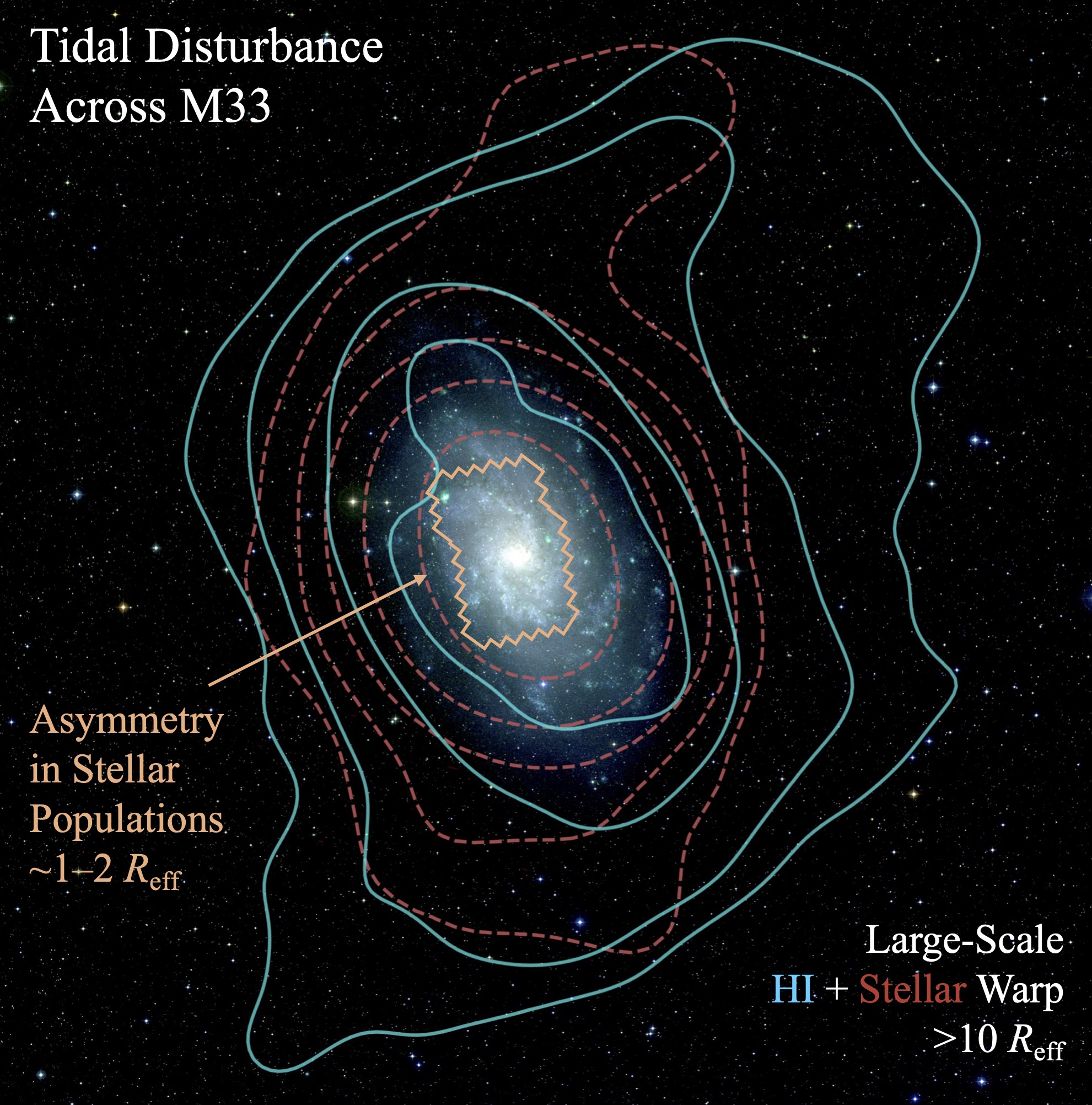}
\caption{A wide-field \textit{BRI} color image of M33, from the Second Generation Digitized Sky Survey (DSS2). The PHATTER footprint is overlaid in orange, showing the scale of the asymmetries in M33's stellar morphology measured in this paper --- evidence of tidal disturbance of M33's inner regions. Smoothed contours of \textsc{H\,i} 21\,cm emission \citep{putman2009} are overlaid in blue, and of metal-poor RGB stars detected by the PAndAS survey \citep{martin2013} overlaid in dashed red, showing the known large-scale gaseous and stellar warp in M33 --- evidence of tidal disturbance of M33's outskirts.}
\label{fig:hi}
\end{figure*}

It has been long known that M33's outer \textsc{H\,i} disk features a prominent warp with respect to its stellar disk \citep[e.g.,][]{rogstad1976,corbelli&edvige1997,putman2009}. We show the \textsc{H\,i} distribution of M33 overlaid on an optical image in Figure \ref{fig:hi}. More recently, a similar warp feature has been discovered in M33's stellar populations at large radii \citep[e.g.,][]{mcconnachie2010,cockroft2013,mcmonigal2016}, with the favored origin the tidal disruption of M33's outskirts due to a recent interaction. We also show the distribution of this large-scale stellar substructure in Figure \ref{fig:hi}, where it shows a similar extent to the H\,\textsc{i}. 

As shown in Figure \ref{fig:hi}, this large-scale warp impacts the galaxy at distances far outside the PHATTER footprint. Yet, our results suggest that the inner parts of M33 have not escaped unscathed. As discussed in \S\,\ref{sec:spiral}, we find that M33's two primary spiral arms, with pitch angles $\sim$24\textdegree, clearly visible in Figures \ref{fig:maps} \&\ \ref{fig:spiral}, are offset from one another by $\sim$15\textdegree\ in phase. Moreover, the strength of these two arms is highly asymmetric, with the southeastern arm exhibiting a $\sim$50\% density enhancement over the northwestern arm. This kind of asymmetry is expected in cases where a tidal interaction has induced additional spiral modes, as has been suggested for M51 \citep{henry2003}. It is also consistent with the results of \cite{koch2018}, who found asymmetry in the H\,\textsc{i} line wings in M33's inner disk thought to be a component of the gas that has been dynamically heated. It is interesting to note that the direction of M33's outer warp opposes the direction of its spiral arms. 

These results suggest that while M33's outskirts have been the most prominently impacted, the \textit{entire disk} has likely been affected by a tidal interaction in the recent past. This deepens an existing mystery: what did M33 recently interact with? An obvious possibility is that M33 experienced a recent close encounter with M31. The classical \textit{S}-shape of its outer stellar and gaseous distributions is typical of `flyby' encounters between the disrupting galaxy and a larger galaxy, in both simulations \citep[e.g.,][]{besla2012,gomez2017,semczuk2020} and observations of other nearby galaxies \citep[e.g., And I, NGC 3077;][]{mcconnachie&irwin2006,okamoto2015,smercina2020}. M31 is the closest galaxy that is more massive than M33, and indeed the tidal structures in M33's H\,\textsc{i} and stars point in the direction of M31. Yet, orbital modeling of M33, using proper motions of water maser emission and stars from VLBA, HST, and Gaia observations \citep{brunthaler2005,sohn2012,vandermarel2012b,vandermarel2019}, strongly suggests that it is on its first infall into the Local Group and has not likely yet experienced a close pericenter passage around M31 \citep{patel2017,vandermarel2019}. M33's clear disk-wide signatures of tidal interaction and orbital parameters inferred from proper motion observations therefore remain in tension. 

In addition to the possibility of `long-range' tidal influence from M31 \citep[e.g.,][]{vandermarel2019}, we offer two additional avenues to settling this tension. First, analysis of the distribution, SFH, and abundance patterns in M31's stellar halo populations provides strong comprehensive evidence that there was an additional `major member' of the Andromeda system only several billion years ago: likely the progenitor of the M31's Giant Stellar Stream (GSS), and possibly also the origin of the compact dwarf elliptical M32 \citep[e.g.,][]{dsouza&bell2018b,hammer2018,escala2021}. By some measures, this galaxy was nearly 1/3 as massive as M31 itself and merged with M31 only $\sim$2\,Gyr ago \citep{dsouza&bell2018b} --- meaning that it is possible that this disrupted merger partner could have interacted with M33 prior to its coalescence with M31. M33's resolved SFH does appear to show a global enhancement within the last 2\,Gyr \citep{williams2009}. While idealized models attempting to explain the morphology of the GSS with a less-massive progenitor have a preference for the progenitor approaching from the northwest \citep[e.g.,][]{kirihara2017}, the orbit of this disrupted massive galaxy is ultimately not well constrained, therefore it is difficult to assess the feasibility of this past interaction. However, observations of M31's stellar halo, such as the recent wide-field spectroscopic observations from the Dark Energy Survey Instrument \citep[DESI;][]{dey2022} and planned observations with the upcoming Subaru Prime Focus Spectrograph, together with insight from dynamical simulations, may shed further light on this possibility. 

Second, we offer that it may be important to consider the impact of this recent massive accretion on the \textit{gravitational potential of M31}, which has not yet been incorporated in studies of M33's orbital history \citep[e.g.,][]{patel2017,vandermarel2019}. Recent work has found that the MW's accretion of the LMC, and the induced dark matter wake, has had an enormous impact on its global potential \citep{garavito-camargo2019,conroy2021}. Building on this, theoretical work has shown that failing to model the potential of the LMC, and its impact on the MW potential, leads to considerable uncertainties on the inferred orbital properties of MW satellites from backwards integration \citep{dsouza&bell2022}. The GSS progenitor was likely substantially more massive than the LMC (in an absolute sense, and relative to M31), which would have resulted in an even larger impact on M31's potential than what has been observed in the MW. In fact, the variance on M31's gravitational potential induced by this merger is likely comparable to current uncertainties on the M31's distance, proper motion, and particularly total mass \citep[e.g.,][]{dsouza&bell2022}. There may, therefore, still be room for dynamical solutions favoring a close passage between M33 and M31 if the impact of this likely massive, recent merger were to be accurately modeled. 

In summary: we argue that the interaction history of M33 in the M31 group is not a settled issue, but rather requires additional observational and theoretical study. Future work on the spatially-resolved ancient SFH of M33 using the PHATTER data will also shed light on its recent interaction history. 

\subsubsection{Stability of M33's Disk}
\label{sec:potential}

The contrast of M33's structure (and that of other nearby galaxies) in the optical and near-infrared has been observed in a number of studies over the years \citep[e.g.,][]{regan&vogel1994,jarrett2003}. Yet, the contrast is particularly stark in this work, where we are able to cleanly separate the young and ancient stellar populations. It is first worth noting that overall intuition about what galaxies \textit{should} look like at this level of fidelity is incredibly limited. It requires resolving stars across the entire disk of a galaxy, limiting us to very nearby examples. Only three relatively undisturbed galaxies that host spiral structure exist within the Local Group: the MW, M31, and M33. Studying the Galaxy's structure from within presents its own challenges. Therefore, some of the surprise over M33's dramatic age-dependent morphology comes from a lack of prior context. Dramatic differences between the distribution of young and old stars in galaxies may be more common than we currently realize.

Furthermore, the typical interpretation of galaxy morphology in the Hubble Diagram has largely been shaped by integrated light imaging of both nearby and more distant galaxies, where many different stellar populations are blended together. Indeed, in star-forming galaxies the brightest stars are young and massive, still cohabitating with the collisional ISM from which they were born, and can strongly influence the galaxy's morphological classification \citep[e.g.,][]{meidt2012}. With that said, a comparative example is the PHAT survey's study of M31. In this galaxy, with an order of magnitude higher stellar mass than M33, the ISM more closely follows the underlying morphology exhibited by the dominant ancient stellar component \citep{dalcanton2015}. M33, in contrast, has a much lower central stellar density than an M31-mass disk galaxy, and its ISM constitutes a much larger fraction of its mass budget \citep[$M_{\rm \textsc{H\,i}}/M_{\star}$\,$>$\,30\% within the PHATTER footprint; e.g.,][]{corbelli2014,sellwood2019}. Both M33's neutral \citep[\textsc{H\,i};][]{koch2018} and molecular \citep[CO;][]{druard2014} ISM exhibit a completely flocculent structure, very similar to the MS stellar density map in Figure \ref{fig:maps}. Why does M33's ISM behave so differently from the bulk of the stars? 

\begin{figure*}[t]
\centering
\leavevmode
\includegraphics[width={\linewidth}]{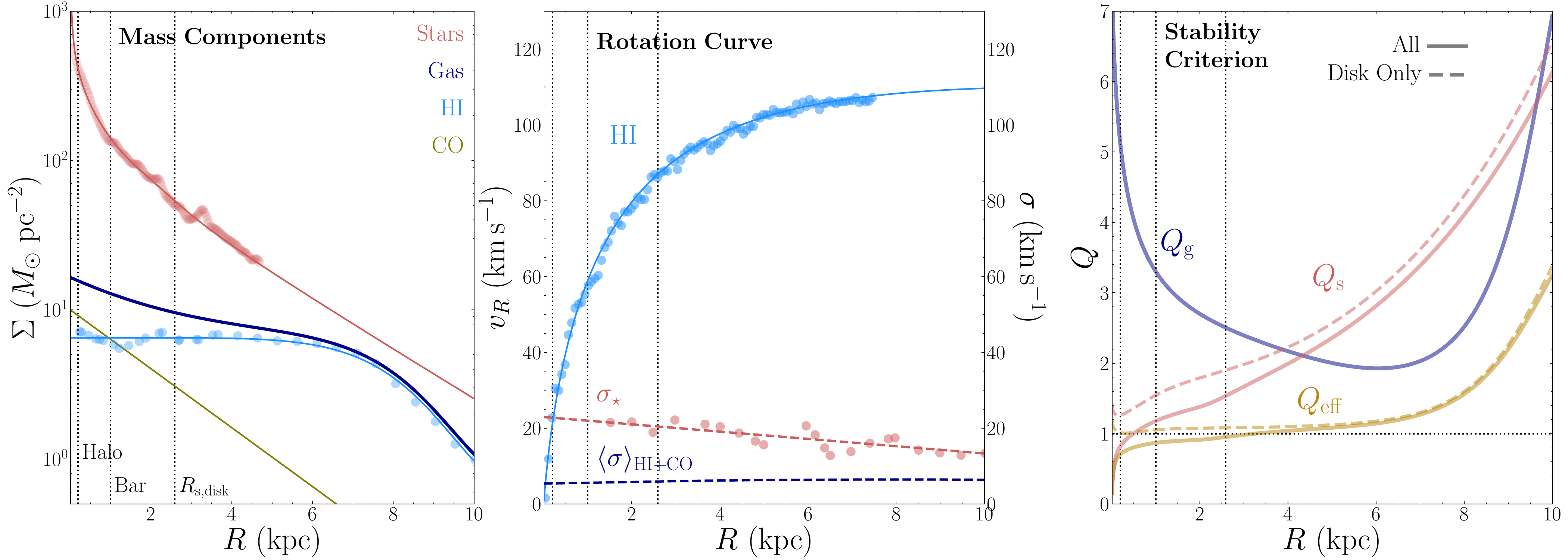}
\caption{Analysis of the global stability of M33's disk. In each panel, the two black dashed lines denote the scale at which the halo and bar components dominate, respectively, as well as the scale length of the disk. \textit{Left}: Surface mass density distributions of M33's primary components, including stellar (this work; red) and gaseous (H\,\textsc{i} and CO; \citealt{corbelli2014}). \textit{Center}: H\,\textsc{i} rotation curve (blue points) with corresponding Brandt rotation curve model fit (blue curve) from \cite{koch2018}, and velocity dispersion profiles of the stars (red) \citep{quirk2022} and gas (blue) \citep{koch2019}. \textit{Right}: Toomre $Q$\ stability criterion for gas, stars, and $Q_{\rm eff}$, calculated from the mass distributions, rotation curve, and velocity dispersion profiles. We show the results for both the full stellar distribution (solid), and only the disk components (the bar and exponential disk; dashed). The canonical criterion for the a `stable' disk is shown as a dotted line ($Q\,{>}\,1$).}
\label{fig:stability}
\end{figure*}

The stability of M33's disk has been a topic of significant recent interest. \cite{sellwood2019}, for example, found it nearly inexplicable that M33 did not possess a bar given the stellar and gaseous content of its disk. In light of recent observational insights into the dynamics of M33's stellar and gaseous components, we re-examine the stability of its disk in the context of the ubiquitous stability criterion, $Q$\ \citep{toomre1964}, for an infinitely thin, self-gravitating disk. $Q$\ is defined as
\begin{equation}
    Q \equiv \frac{\sigma_r \kappa}{\pi G \Sigma}, 
\end{equation}
where the epicyclic frequency, $\kappa$, is given by
\begin{equation}
    \kappa = \sqrt{2} \bigg(\frac{v_r^2}{r^2} + \frac{v_r}{r} \frac{dv_r}{dr}\bigg)^{1/2}. 
\end{equation}
Figure \ref{fig:stability} shows the results of this analysis. For the stellar mass density profile, we adopt the RGB density profile measured in this work, normalized to a total stellar mass of $3{\times}10^9\,M_{\odot}$. The profiles for M33's atomic and molecular gas we adopt from \cite{corbelli2014}, and use the updated H\,\textsc{i} rotation curve from \cite{koch2018}. Recently, the velocity dispersion profiles of the stars \citep{quirk2022} and gas \citep{koch2019}\footnote{The gas velocity dispersion profile is calculated as a mass-weighted average from the H\,\textsc{i} and CO mass surface density profiles (Figure \ref{fig:stability}, left panel). In contrast, many H\,\textsc{i} surveys assume a fixed velocity dispersion of 10\,km\,s$^{-1}$\ for the neutral gas \citep[e.g.,][]{tamburro2009}. If we assume a similar, slightly higher gas dispersion, it does not change the qualitative results of our analysis.} have been measured, allowing a direct calculation of $Q$. We adopt the common \cite{wang&silk1994} approximation for an ``effective'' stability criterion\footnote{For illustrative purposes, we adopt the \cite{wang&silk1994} approximation over more detailed solutions \citep[e.g.,][]{rafikov2001,romeo&wiegart2011} for simplicity.} in the case of a mass distribution of gas and stars: 
\begin{equation}
Q_{\rm eff}^{-1} = Q_{\rm g}^{-1} + Q_{\rm s}^{-1}
\end{equation}
Given that the halo component we infer is likely spheroidal accounts for substantial mass at M33's center, and is not included in the velocity dispersion profile presented in \cite{quirk2022}, its impact on the stability criterion is somewhat unclear. We therefore consider both the cases: the full stellar mass profile and the case of just the disk components, the exponential disk and the bar. We note that this has only a modest effect on the innermost values of $Q_{\rm s}$\ and $Q_{\rm eff}$. 

Overall, we find that within the inner 1--2 effective radii, M33's disk is very close to the stability threshold of $Q\,{=}\,1$\ (Figure \ref{fig:stability}, right). This is similar to studies of M33 using hydrodynamical simulations that found M33's disk to be only marginally stable \citep{dobbs2018,sellwood2019}. Our results support this conclusion, as Figure \ref{fig:stability} shows that M33's disk is not unusually stable, with $Q\,{\sim}\,1$\ across nearly the entire disk, out to $\sim$3 disk scale lengths. Its barred, two-arm spiral structure is therefore well in line with expectations given the mass of its disk. However, the question of why the gas (and young stars) and bulk stellar mass are ordered so differently remains. A simple answer is that the vastly smaller characteristic scales of gravitational instabilities in the gas and collisional nature, compared to the stars, invites more fragmented structure in the ISM, as given by $\lambda_{\rm crit}\,{=}\,4\pi^2 G \Sigma/\kappa^2$\ \citep{toomre1964} and seen in isolated simulations of low-mass galaxy disks \citep[e.g.,][]{dobbs2018,sellwood2019}. However, this should be the case for all similarly gas-poor galaxies, regardless of overall mass. Indeed, images of nearby galaxies in the ultraviolet or H$\alpha$\ show a much greater degree of structure compared to, for example, the optical or near-infrared \citep[e.g.,][and many others]{dale2009}. Why does M33's composite visual structure, in particular, differ so substantially from the morphology of its bulk stellar mass? 

While a dynamically-motivated answer to this question is beyond our ability to infer from these observations, important observational context can be found in other nearby galaxies. First, it has become clear empirically that the typical structure of the ISM in spiral galaxies is highly dependent on stellar mass. \cite{davis2022}, for example, recently showed that the molecular ISM of galaxies becomes increasingly flocculent as their effective stellar surface density decreases. M33 follows this trend, and so, empirically, the flocculence of its young stars should be unsurprising. Further, while very few studies of the global stellar populations of nearby galaxies have been conducted to a similar depth as PHATTER, there is subtle existing evidence that lower-mass disk galaxies exhibit increasingly different morphologies depending on the wavelength at which they are observed. For example, as part of the S$^4$G Survey, \cite{buta2015} found that for late-type spirals classified Scd\footnote{M33 is classified as an Scd spiral in optical images \citep{deVaucouleurs1991}.} to Sm in optical images (i.e., flocculent spirals), the most common classification at 3.6\,\micron\ was SB. The presumption in that study was that this was due to the near-infrared more closely tracing the bulk of the stellar mass, rather than the young, low mass-to-light populations that are prominent in $B$-band. With access to M33's true stellar populations across the disk, we agree with this conclusion, though a theoretical understanding of this increase in flocculent structure in lower-mass disks is still needed. Though its bar appears to be more prominent overall, it is interesting to note that the LMC's bar also appears to disappear when considering only the distribution of young, luminous stellar populations, compared to the ancient populations \citep[e.g.,][]{kim2003,choi2018b}. 

\begin{figure*}[t]
\centering
\leavevmode
\includegraphics[width={0.8\linewidth}]{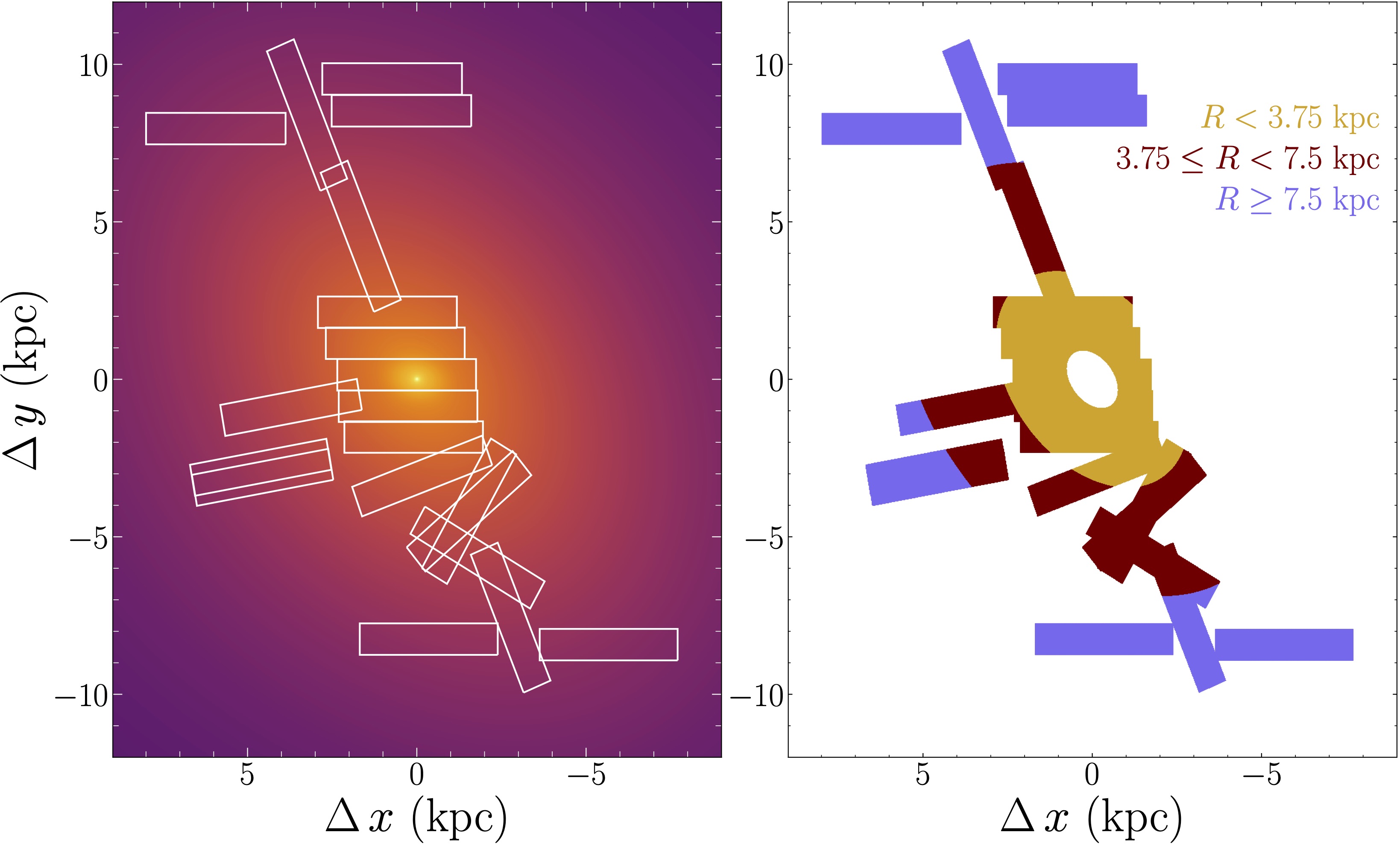}
\caption{\textit{Left}: A 2D model constructed from the three best-fit components found in our density profile decomposition, in units of stellar density and displayed using a logarithmic scale. The positions of the Keck DEIMOS slitmasks used in the TREX survey \citep{gilbert2022,quirk2022} are shown as white rectangles. \textit{Right}: The outlines of the DEIMOS slitmasks, color-coded to the three bins of radius defined in \cite{gilbert2022} for measuring the mass fraction of the non-rotating stellar halo component. The innermost $\sim$1\,kpc is masked, as \cite{gilbert2022} did not report average velocities at these innermost radii.}
\label{fig:trex}
\end{figure*}

In summary, the flocculent nature of M33's ISM, which is in line with other similarly low-mass disks, naturally results in a flocculent structure for the youngest populations. These young, low mass-to-light populations empirically appear to bias morphological inferences of the underlying bulk stellar populations, particularly in lower-mass disks. The `hidden' barred, two-arm structure in M33 brings our observational picture of this well-studied nearby galaxy into harmony with theoretical predictions \citep[e.g.,][]{sellwood2019}. As this explanation is not dependent on M33's past or current state of interaction, it is feasible that it could also be relevant for M33-analogs in the Local Volume. Could, for example, galaxies such as NGC 300, NGC 2403, or NGC 2976 similarly be hiding barred, grand design spirals beneath the flocculent structure of their youngest, most luminous stellar populations? Future global studies of their global resolved stellar populations could help us understand similarities or differences in their structure compared to M33. 

\subsection{The Stellar Halo of M33}
\label{sec:halo}

The search for a stellar halo around M33 is well into its fourth decade \citep[e.g.,][]{mould&kristian1986,mcconnachie2006,mcconnachie2010,mcmonigal2016,gilbert2022}. It appears that M33's stellar outskirts are likely dominated by disk stars tidally disrupted in some past interaction with an unknown perturber (\citealt{mcconnachie2010}, and see discussion in \S\,\ref{sec:interaction}). Yet, evidence for M33's own accreted stellar halo via star counts has remained elusive \citep[e.g.,][]{mcmonigal2016}. Excitingly, recent spectroscopic observations of RGB stars identified via HST photometry from PHATTER, and ground-based photometry from the PAndAS survey, revealed a distinct high velocity dispersion, non-rotating stellar component in M33 \citep[the TREX Survey;][]{gilbert2022,quirk2022} --- consistent with a stellar halo population. 

\begin{figure*}[t]
\centering
\leavevmode
\includegraphics[width={\linewidth}]{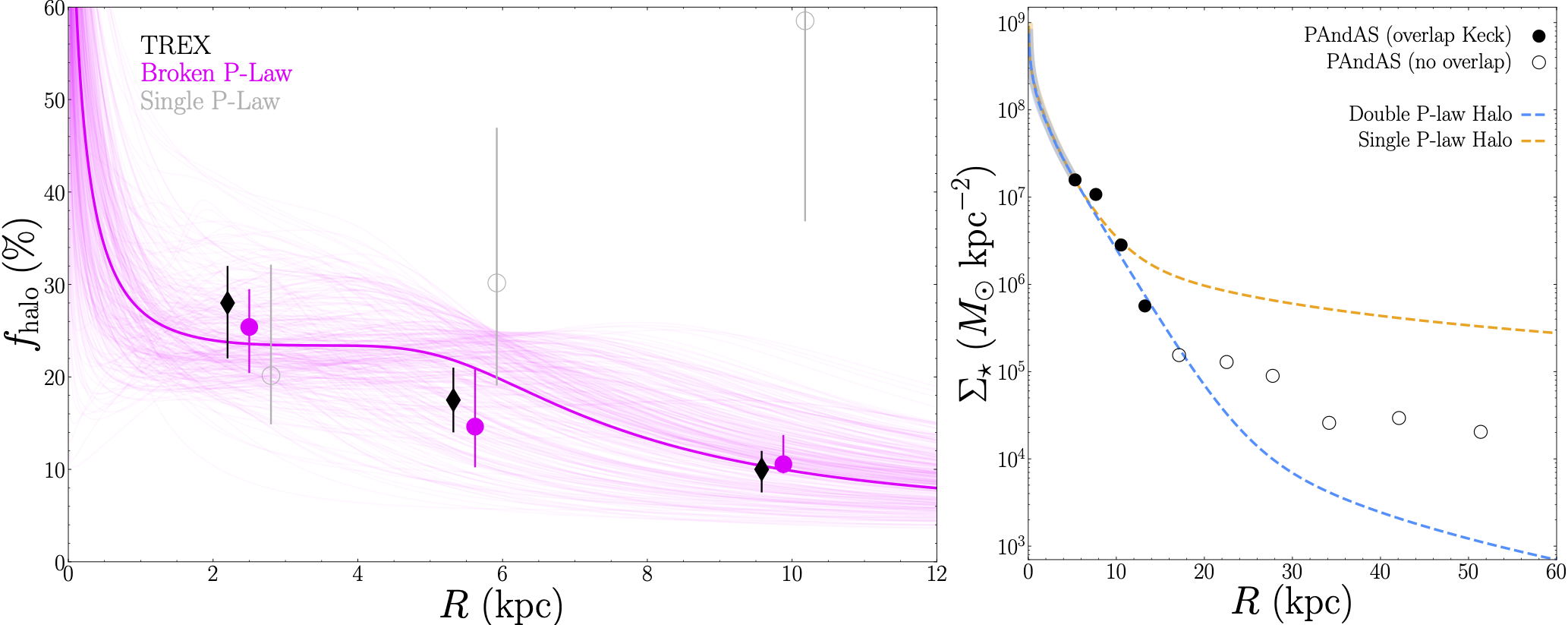}
\caption{Consistency of our broken power-law halo model with evidence for M33's accreted stellar halo within the disk and in its outskirts. \textit{Left}: The global fraction of the stellar halo component, as a function of radius, for our best-fit model, shown as a pink curve. We also show 300 randomly-sampled realizations from the MCMC analysis as transparent pink curves. We show the halo fractions estimated from the TREX survey, in their three broad radial bins, as black diamonds. We also show the average halo fractions in these radial bins, measured within the DEIMOS slitmasks shown in Figure \ref{fig:trex}, for the broken power-law (pink circles) and single power-law halo models (gray circles), with error bars encompassing the 16--84\% range of MCMC realizations. As is clearly visible, the broken power-law halo model matches well reproduces the TREX halo fraction estimates. \textit{Right}: Our best-fit density profile model, assuming both a broken power-law (blue) and single power-law (orange) halo, converted to stellar mass surface density using a 10\,Gyr, [M/H]\,$=$\,$-$1.3 isochrone stellar model. The portion of the profiles directly constrained by PHATTER ($R\,{\lesssim}\,5$\,kpc) are demarcated by broad shaded lines. Overlaid is the (deprojected) radial density of RGB stars to the east of M33, calculated by the PAndAS survey \citep{mcconnachie2010}, and scaled to the density profile measured in this work, for the single point that overlaps in radius. We demarcate the PAndAS points based on whether they overlap with the Keck observations from TREX (filled black) or not (black open). The power-law halo models are therefore \textit{directly} constrained, either by star counts or the halo fractions from Keck spectroscopy, out to $\sim$13\,kpc, and is shown as an extrapolation beyond this point. }
\label{fig:halo}
\end{figure*}

While \cite{gilbert2022} did not measure a radial profile for the halo component, they found that the mass fraction of the halo component declines with radius across three broad radial bins. In Figure \ref{fig:trex}, we show the positions of the TREX slitmasks (left) overlaid on a smooth 2D model image of stellar density, constructed from the best-fit morphological components to our RGB density profile (Figure \ref{fig:profile}), and the corresponding radial bins defined in \cite{gilbert2022} for stars within those masks (right). 

We use the regions shown in Figure \ref{fig:trex} to calculate the stellar halo fraction of our star-count based halo model in each of these three radial bins from \cite{gilbert2022}. In Figure \ref{fig:halo} we show the global halo mass fraction as a function of radius for our full three-component model (pink curve), as well as the averages within the broad radial bins measured by TREX (pink points). For comparison, we also show the halo fractions\footnote{\cite{gilbert2022} did not convert star counts to stellar mass, but the ratio of RGB stars identified in each population is directly analogous to a stellar mass fraction in this case.} measured in the three TREX radial bins for a best-fit model with a single power-law halo component (see \textsc{Appendix} \ref{sec:singleplaw}). This figure shows clearly that our recovered broken power-law model reproduces almost perfectly the halo mass fractions estimated by the TREX survey. In contrast to the broken power-law fit, the best-fit single power-law model exhibits entirely different behavior, with the halo fraction increasing with radius. This is expected, as without a break, the exponential disk declines faster than a single power-law that can fit the central upturn in M33's density profile. 

To compare this halo model with previous studies, it is useful to convert this halo model to more general physical units, such as stellar mass density or surface brightness. To do so, we assume an ancient (10\,Gyr) stellar population for the halo component, with a metallicity of [M/H]\,$=$\,$-$1.3 --- the peak of the metallicity distribution function for the high-dispersion component in \cite{gilbert2022}. Using a 10\,Gyr, [M/H]\,$=$\,$-$1.3 isochrone \citep[from the PARSEC models; e.g.,][]{bressan2012}, and assuming the depth of our RGB selection shown in Figure \ref{fig:selection}, we then estimate a mass conversion of 3036\,$M_{\odot}$ per RGB star. Using this conversion, in the right panel of Figure \ref{fig:halo} we plot our best-fit halo model, converted from star counts to stellar mass surface density, compared to the eastern `off-stream' radial star-count profile calculated by \cite{mcconnachie2010}, which at large radii was argued to be dominated by M33's diffuse stellar halo. As the \cite{mcconnachie2010} profile is based on a very different photometric catalog (e.g., PAndAS; see also \citealt{ibata2014}), we adopt a simple scaling to put the two on the same footing, by matching the PAndAS profile, in RGB counts per degree$^2$, to the global PHATTER profile, in RGB counts per arcsec$^2$, using the innermost point of the PAndAS profile that overlaps with the profile presented in this work  --- $\sim$4\,kpc. This radius is close enough to M33's center that it is reliably disk-dominated, providing an independent comparison between our halo-only model and the PAndAS profile at large radii. 

Our model with the broken power-law halo matches the scaled PAndAS density profile quite well out to 20\,kpc. After this, the PAndAS profile follows a similar radial trend (power-law index of ${\sim}\,{-}2.8$ vs.\ ${-}3.1$\ for the model), but at densities ${\sim}10{\times}$\ higher than we predict. This difference is consistent with unsubtracted background contamination, which is not surprising as it has been argued that the PAndAS profile is still contaminated by MW foreground stars at this level \citep{cockroft2013,mcmonigal2016}. In fact, the difference between these two profiles could be explained by a single RGB star in a 10\,arcmin$^2$\ region --- approximately the size of a single HST ACS field. It therefore seems likely that the PAndAS observations are consistent with the predicted halo model, given the higher purity and depth of the PHATTER RGB sample. Normalizing to M33's estimated global stellar mass, $M_{\star}\,{=}\,3{\times}10^9\,M_{\odot}$\ \citep{vandermarel2012b}, we integrate our broken power-law halo model and predict a total mass of $M_{\rm \star,halo}\,{\sim}\,5{\times}10^8\,M_{\odot}$, the majority of which resides within the central 2.3\,kpc and is well-constrained by both the PHATTER star counts and Keck-sourced stellar kinematics --- a global `halo fraction' of $f_{\rm halo}\,{=}$\,15.6\%. We estimate that only $\sim$8\% of the halo mass resides at $R\,{>}\,13$\,kpc, the regime used by previous studies to place upper limits on M33's halo at ${\lesssim}10^6\,M_{\odot}$\ \citep[e.g.,][]{cockroft2013,mcmonigal2016}, and which is unconstrained by either PHATTER or the existing Keck spectroscopy. We predict a factor of ten higher mass at these radii, $\sim$a few times $10^7\,M_{\odot}$\ (factoring in differences in the agreed-upon total stellar mass of M33). Given the extremely low star counts in PAndAS at these radii, and the independent consistency between our halo model and both the TREX kinematics the PAndAS data within 20\,kpc, it seems likely that these profiles were slightly oversubtracted, leading to a nondetection of the very faint halo. 

Our results therefore seem unambiguous: we have finally obtained the long-sought-after observationally-motivated model of the stellar halo of M33. The predicted mass (${\sim}\,5{\times}10^8\,M_{\odot}$) and photometrically-inferred metallicity ([Fe/H]\,$\sim$\,$-$1.3; \citealt{gilbert2022}) of M33's stellar halo are completely consistent with the stellar accreted mass--metallicity relationship for nearby MW-mass galaxies \citep{harmsen2017,dsouza&bell2018a,smercina2022}, strongly supporting an accreted origin. Assuming that the origin of M33's stellar halo is accreted, its halo fraction is relatively high, but well within the range for more massive nearby galaxies \citep{harmsen2017}. In fact, if the Magellanic Clouds, which likely interacted in the past \citep[e.g.,][]{besla2012,massana2022}, were not currently interacting with the Milky Way and we could view them in several Gyrs following their final merging, the remnant LMC's `accreted halo fraction' would be nearly identical to what we estimate for M33. 

In follow-up work using the PHATTER catalog, and related datasets, we will seek to further constrain M33's halo using low metallicity-sensitive stellar populations, such as horizontal branch stars, that are more numerous than RGB stars, less likely to be contaminated by MW foreground, and are reliable tracers of metal-poor accreted halos \citep[e.g.,][]{williams2012,williams2015b}.

\section{Conclusions}
\label{sec:conclusions}

In this paper, we have used resolved stellar photometry from the PHATTER survey to characterize the global structure of M31's largest satellite and one of the most massive galaxies in the Local Group --- the Triangulum Galaxy, M33. Using artificial stellar populations, we have selected four distinct populations in color--magnitude space, evenly logarithmically-spaced in stellar age --- the young upper main sequence (MS), core helium-burning/red supergiant (HeB), asymptotic giant branch (AGB), and red giant branch (RGB) stars --- and analyzed M33's structure in each. Using RGB stars, which trace the bulk of M33's stellar mass, we fit elliptical isophotes and decompose the resulting radial profile. From this analysis, we find: 
\begin{enumerate}[topsep=5pt,itemsep=0pt,left=2pt]
    \item The flocculent structure with which M33 is typically associated is only observed in the youngest populations, traced by upper MS and HeB stars, as expected from the structure of its ISM.
    \item The older populations in M33, traced by AGB and RGB stars, are dominated by a smoother, barred disk, with only two visible spiral arms.
    \item The two primary spiral arms are asymmetric in both phase, separated by 165\textdegree, and strength, with the southeastern arm $\sim$50\% stronger than the northwestern arm. We suggest that this asymmetry is likely the innermost evidence of the tidal interaction that formed M33's prominent \textsc{H\,i} and stellar warp at large radii. 
    \item From the decomposition of the radial RGB density profile, the smooth structure of M33 is well-reproduced by the inclusion of a S\'{e}rsic bar with an effective radius of 0.9\,kpc, an accreted stellar halo component characterized by a broken power-law, with an outer slope of $-$3.07, and an exponential disk. 
    \item Using isophotal metrics for bar strength from the literature, we the find M33's bar to be on the low end of classically `strong' bar classifications, but entirely consistent with the range of nearby barred galaxies, with $f_{\rm bar}\,{\sim}\,0.18$. M33 is, indeed, a `hidden' barred galaxy.
    \item Using our census of stellar mass in M33, and taking advantage of recent advances in dynamics of M33's stellar and gaseous components, we find that its disk is only very marginally stable, with $Q\,{\sim}\,1$\ over the majority of its disk. This is very close to the limit of self-regulation and, following recent theoretical studies of M33, supports its formation of a two-arm spiral and central bar. 
    \item Our broken power-law model for M33's stellar halo is highly consistent with both recently-measured fractions of a non-rotating, high velocity dispersion `stellar halo' component in the disk, and with the radial profile of RGB stars at large radius along M33's minor axis. From this model, we predict a total stellar halo mass of ${\sim}5{\times}10^8\,M_{\odot}$, the majority of which resides in the inner 2.3\,kpc. The mass and metallicity of M33's stellar halo are highly consistent with the accreted mass--metallicity relation for nearby galaxies, favoring an accreted origin. The resulting predicted accreted halo mass fraction of $f_{\rm halo}\,{=}$\,15.6\% is entirely consistent with the range observed for more massive nearby galaxies. The predicted outer slope for M33's halo of $-$3.07 beyond 2\,kpc should be useful in future searches for M33's diffuse stellar halo at large radii. 
\end{enumerate}

These results constitute a significant step forward in understanding the structure of the nearest late-type spiral galaxy. The insight gained into the different morphologies of M33's young and ancient stellar populations may serve as an important empirical benchmark for ideas about how low-mass disk galaxies form and evolve. Future studies on the resolved stellar populations in M33, including its global ancient star formation history, will use use the findings presented here to further elucidate M33's evolution in the Local Group and its interaction history. \\

\vspace{13pt}
We thank the anonymous referee for a thoughtful review that improved this paper. This research is based on observations made with the NASA/ESA Hubble Space Telescope obtained from the Space Telescope Science Institute, which is operated by the Association of Universities for Research in Astronomy, Inc., under NASA contract NAS 5–26555. 

A.S.\ and M.D. were supported by NASA through grant \#GO-14610 from the Space Telescope Science Institute. E.F.B.\ was partly supported by the National Science Foundation through grant 2007065 and by the WFIRST Infrared Nearby Galaxies Survey (WINGS) collaboration through NASA grant NNG16PJ28C through subcontract from the University of Washington. E.W.K acknowledges support from the Smithsonian Institution as a Submillimeter Array (SMA) Fellow and the Natural Sciences and Engineering Research Council of Canada. The Flatiron Institute is funded by the Simons Foundation.

The Digitized Sky Surveys were produced at the Space Telescope Science Institute under U.S. Government grant NAG W-2166. The images of these surveys are based on photographic data obtained using the Oschin Schmidt Telescope on Palomar Mountain and the UK Schmidt Telescope. The plates were processed into the present compressed digital form with the permission of these institutions.

\facility{Hubble Space Telescope, Spitzer Space Telescope, Keck Observatory, CTIO, IRSA}

\software{\texttt{Matplotlib} \citep{matplotlib}, \texttt{NumPy} \citep{numpy-guide,harris2020array}, \texttt{Astropy} \citep{astropy}, \texttt{SciPy} \citep{scipy}, \texttt{PhotUtils} \citep{photutils}, \texttt{emcee} \citep{foreman-mackey2013}, \texttt{corner} \citep{corner}}

\bibliographystyle{aasjournal}
\bibliography{references}

\appendix

\section{Ages of Synthetic Populations from Alternative SFHs}
\label{sec:alt-sfhs}

\begin{figure*}[t]
\centering
\leavevmode
\includegraphics[width={0.97\linewidth}]{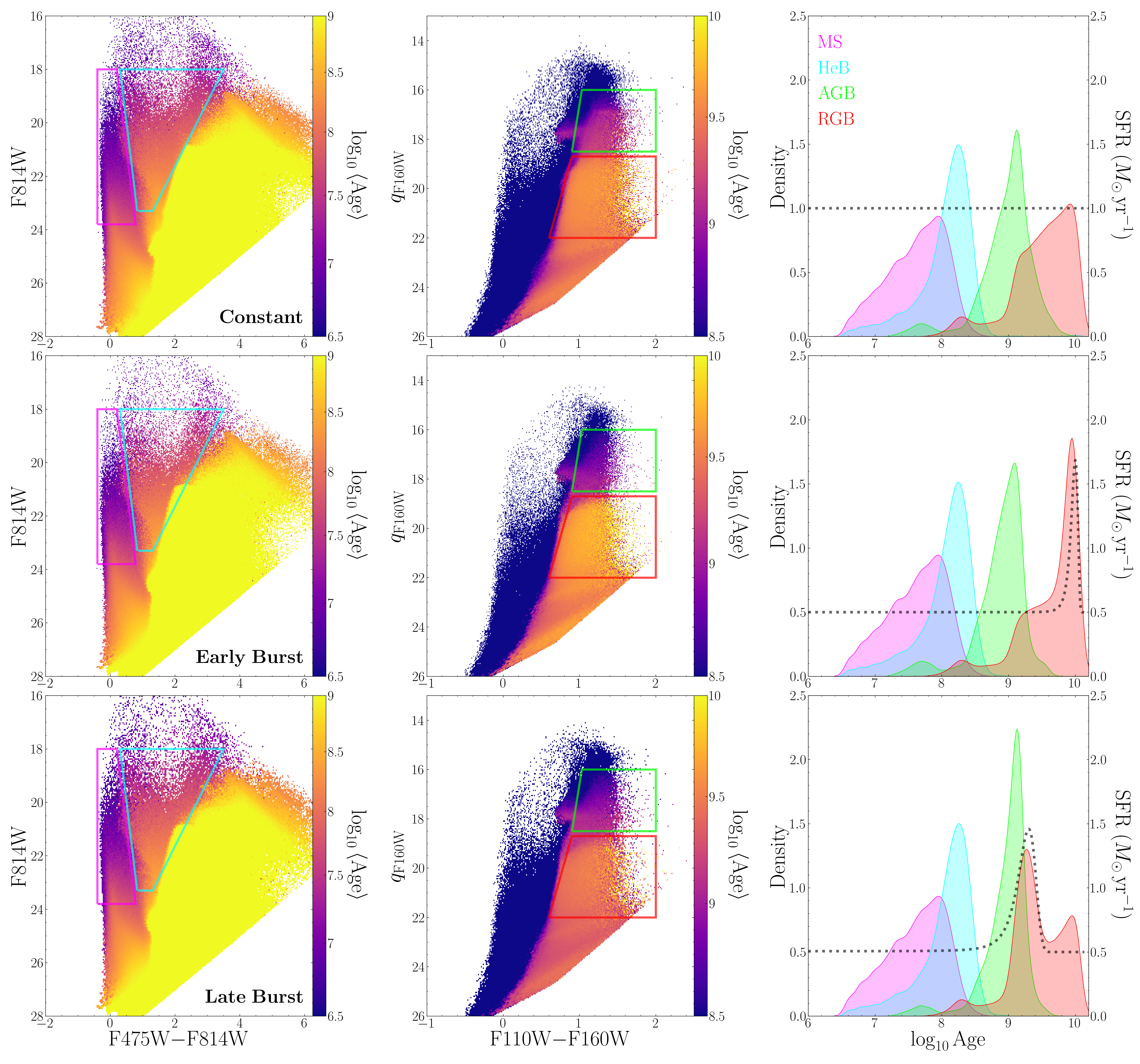}
\caption{Same as Figure \ref{fig:fakestars}, but showing the three cases described in Appendix \ref{sec:alt-sfhs}: an early burst peaking at 10\,Gyr and a late burst peaking at 2\,Gyr, each of which formed half of the stars. The SFH for each case is shown as a dotted black line in the right panel. As is visually apparent, the RGB is, by far, the most sensitive to these different SFHs.}
\label{fig:alt-fakestars}
\vspace{17pt}
\end{figure*}

In this section, we consider two additional cases for the SFH of the synthetic populations we use to estimate the approximate age distributions of the four populations considered in this work, described in \S\,\ref{sec:alt-sfhs}. The global recent SFH of M33 is well constrained by \cite{lazzarini2022}, showing that it is relatively well approximated as a constant SFH over at least the past $\sim$600\,Myr. To assess potential uncertainties on our reported ages for the older populations, we consider two additional limiting cases: a strong burst at very early times, peaking at an age of 10\,Gyr, and a much more recent burst peaking at an age of 2\,Gyr. We model these bursts as the sum of a constant SF component and a log-normal burst \citep{gladders2013,diemer2017}, such that each burst lasts for a total duration of $\sim$2\,Gyr and forms 50\% of the total formed stellar mass of the synthetic population. We show the resulting age-coded Hess diagrams and population age distributions for these different SFHs, including the constant SFH case adopted throughout the rest of the paper, in Figure \ref{fig:alt-fakestars}.

As expected, the young star age distribution is nearly identical to the purely constant SFH case. The HeB and AGB selections show modest deviations, but are very similar to the constant SFH case. The RGB selection shows the largest deviation, as expected, and the shape of its age distribution is sensitive to the shape of the SFH. However, the age hierarchy of these selections remains the same: the MS traces the youngest, most luminous stars, the HeB traces the slightly older populations with ages of several hundred Myrs, the AGB traces the evolved populations on the order of $\sim$1\,Gyr, and the RGB stars trace the truly ancient populations on the order of a few-to-many Gyrs. We note that there is additional uncertainty inherent in our assumption of a uniform age--metallicity relation for all SFH cases. While this is the best that can be done currently, and indeed assuming a well-measured relation from stellar clusters such as the \cite{beasley2015} result improves on the typical age--metallicity degeneracy considerably, it is a limitation that readers should be aware of.

\vspace{-5pt}
\begin{deluxetable}{qssssss}[!ht]
\tablecaption{\textnormal{Population Ages for Different SFHs}\label{tab:ages-appendix}}
\tablecolumns{7}
\setlength{\extrarowheight}{2pt}
\tabletypesize{\small}
\tablehead{%
\colhead{} &
\multicolumn{3}{c}{Early Burst} &
\multicolumn{3}{c}{Late Burst} \\ \cline{2-4} \cline{5-7}
\colhead{} &
\colhead{16\%} &
\colhead{50\%} &
\colhead{84\%} &
\colhead{16\%} &
\colhead{50\%} &
\colhead{84\%} \vspace{-2.5mm} \\
\colhead{Population} & 
\colhead{(Gyr)} & 
\colhead{(Gyr)} & 
\colhead{(Gyr)} \vspace{-5mm}\\ 
}
\startdata
MS & 0.014 & 0.050 & 0.112 & 0.013 & 0.050 & 0.112 \\
HeB & 0.050 & 0.141 & 0.224 & 0.050 & 0.141 & 0.251 \\
AGB & 0.40 & 0.89 & 1.42 & 0.50 & 1.12 & 1.59 \\
RGB & 1.58 & 6.31 & 10.00 & 1.26 & 2.24 & 7.95 \\
\enddata
\tablecomments{Same as Table \ref{tab:ages} in \S\,\ref{sec:fakepop}, but for two additional SFH cases, an early (peak = 10\,Gyr) and late (peak = 2\,Gyr) burst of star formation that each formed half the stellar mass. Median ages and 16--84\% ranges are reported for each of the four different population selections. The values listed for the RGB are for the `inner' (higher-density) RGB selection, which is uniform across the disk.}
\end{deluxetable}

\vspace{-35pt}
\section{MCMC Results for Single and Broken Power Law Halo Component fits}
\label{sec:singleplaw}

\begin{figure*}[t]
\centering
\leavevmode
\includegraphics[width={\linewidth}]{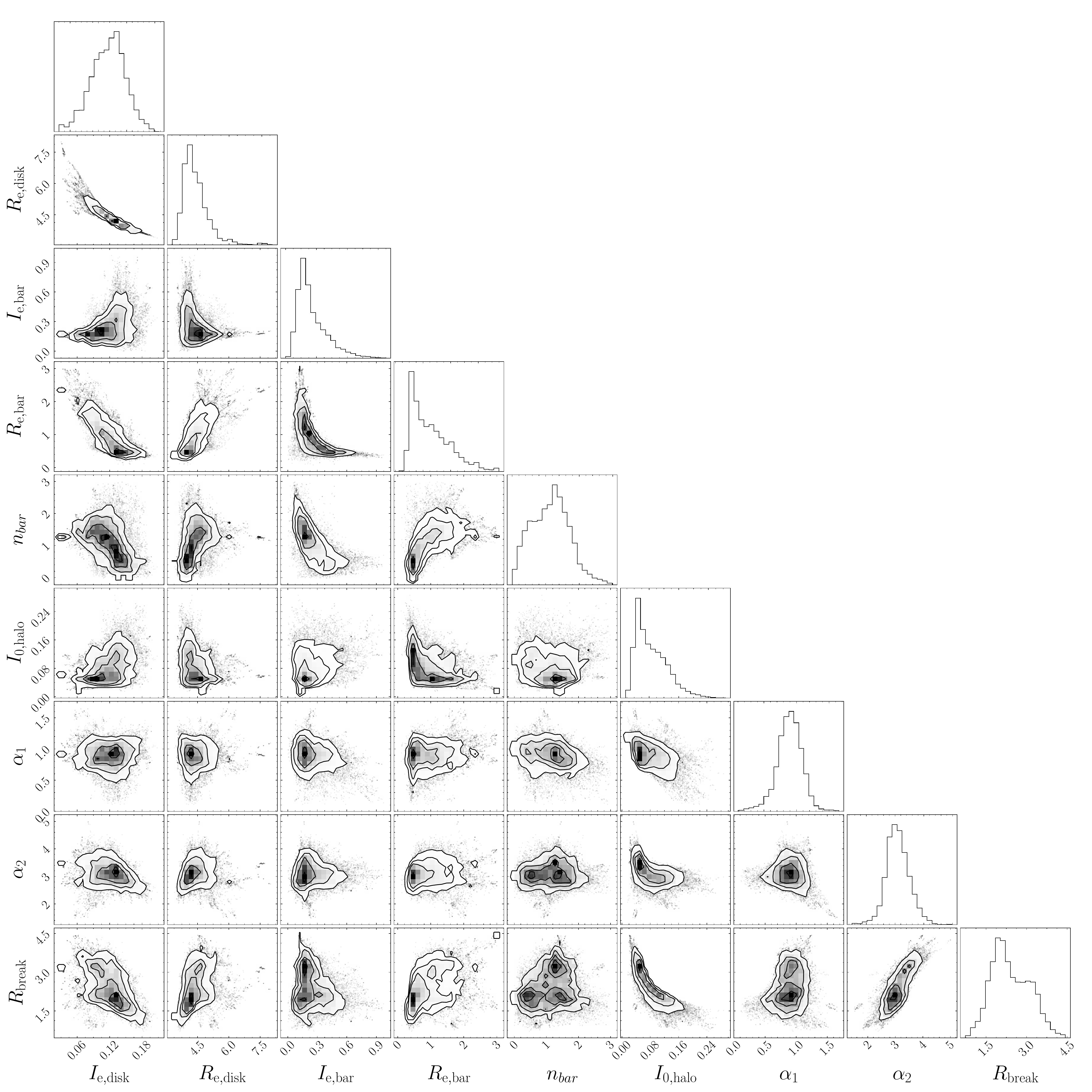}
\caption{\texttt{corner} plot of the single and joint posterior distributions for the nine parameters included in our MCMC modeling of M33's density profile using \texttt{emcee}.}
\label{fig:corner}
\end{figure*}

In this section we present the full results of our MCMC decomposition of M33's radial density profile. When constructing our likelihood function, a flat prior was assumed for each of our nine parameters in our model, described in \S\,\ref{sec:decomp}. We then ran \texttt{emcee} on our model, using 100 MCMC walkers, for 3000 steps.  Figure \ref{fig:corner} shows the posterior single and posterior distributions extracted by \texttt{emcee}, displayed using the \texttt{corner} package \citep{corner}. The median values for each parameter, and their 16--84\% confidence ranges, are given in Table \ref{tab:fit}.

For comparison, we conducted an identical decomposition of our computed RGB density profile assuming a model with a single power-law of the form: 
\begin{equation}
    \Sigma_{\rm halo}(r) = \Sigma_{\rm 0,halo}\,\big(r\big)^{\alpha}
\end{equation}
While the single and broken power law models yield significant differences at large radii, they do not give substantially different results for the effective radius of the disk: 4.34\,kpc with the double power-law and 3.93\,kpc with the single power-law. This explains why a double power-law model has not been used for M33 previously, as without the spectroscopic evidence at larger radii \citep{gilbert2022}, the two models are largely degenerate at the scale of M33's inner disk. 

It is worth noting that this is true of the addition of a halo component in general: it is a nearly completely degenerate component in lieu of a strong prior on the presence of a bar and a hot `halo' component. When only considering the inner regions, a double S\'{e}rsic model is an equally ``good'' fit to the density profile, and if penalizing the number of model components without this prior, using for example an information criterion, a two-component  model is preferred. However, in this case the exponential-dominated profile is in complete tension with the hot kinematic component at large radii. Therefore, when including the substantial halo fraction prior at large radius, the inclusion of a third, halo component is favored.

\begin{figure*}[t]
\leavevmode
\begin{minipage}{\linewidth}
    \centering
    \includegraphics[width={0.8\linewidth}]{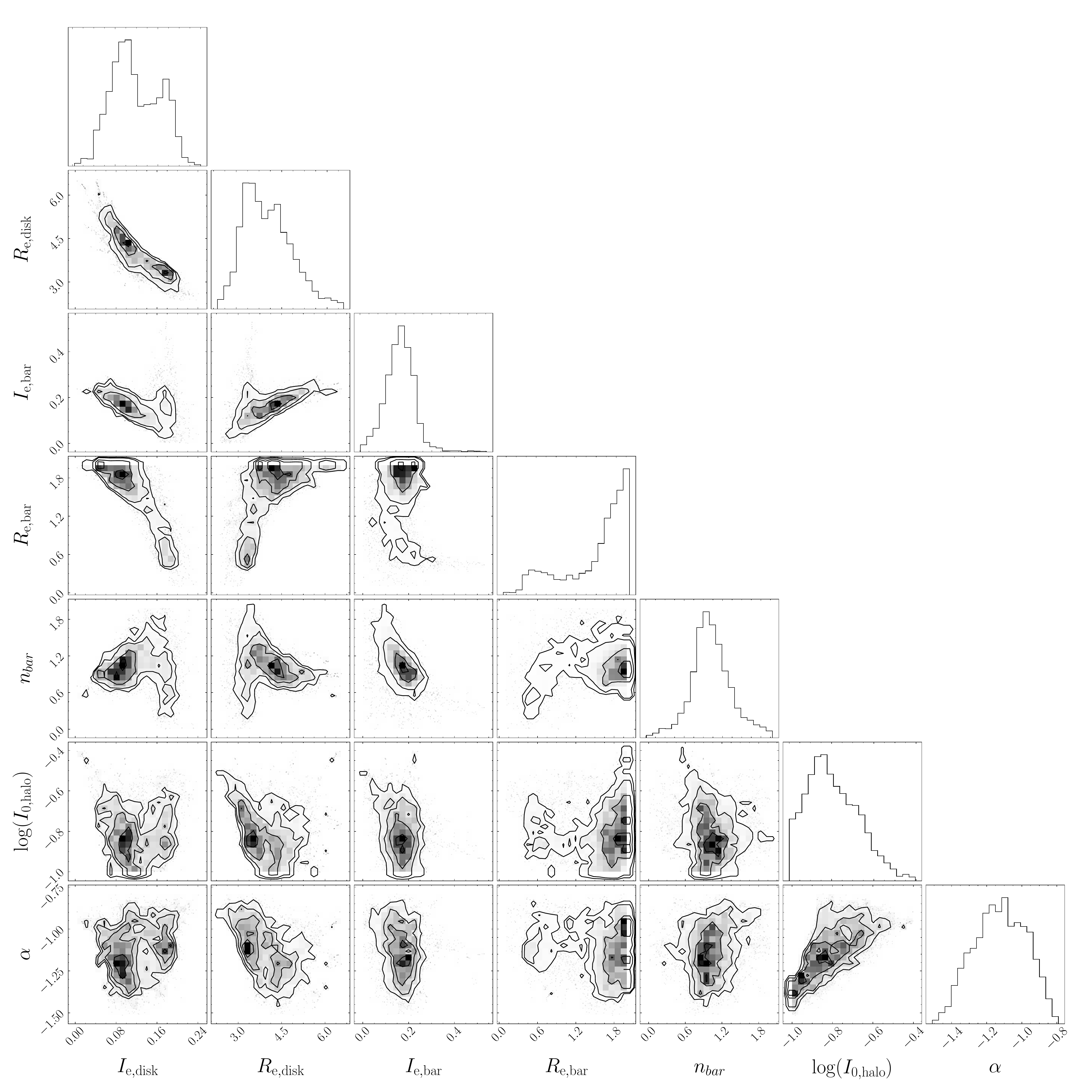}
\end{minipage}
\begin{minipage}{\linewidth}
    \includegraphics[width={\linewidth}]{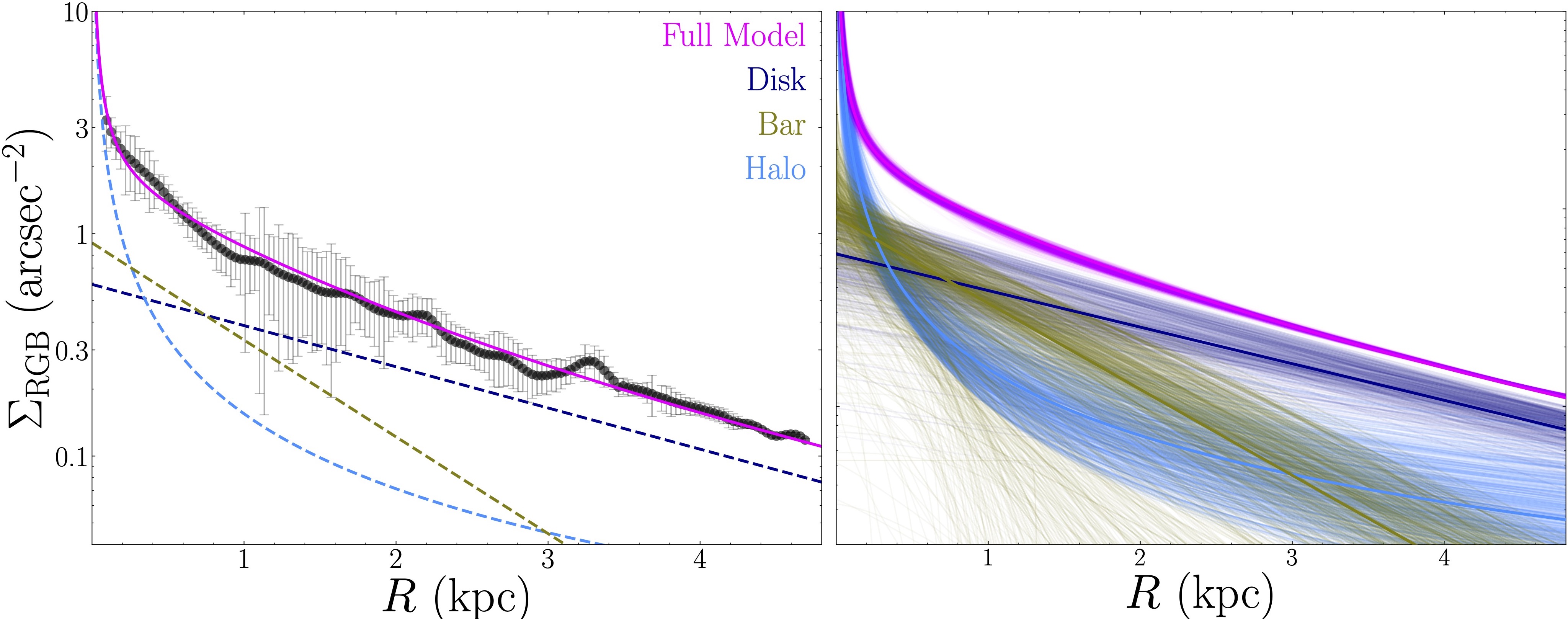}
\end{minipage}
\caption{\textit{Top}: Same as Figure \ref{fig:corner}, but for the case of a six-parameter model with a single power-law parameterization for the stellar halo, rather than the broken power-law adopted throughout the paper. \textit{Bottom}: Same as Figure \ref{fig:profile}, but for the case of a single power-law component for the halo.}
\label{fig:corner-single}
\end{figure*}

\end{document}